

\font\titlefont = cmr10 scaled\magstep 4
\font\sectionfont = cmr10
\font\littlefont = cmr5
\font\eightrm = cmr8 

\def\sss{\scriptscriptstyle} 

\magnification = 1200

\global\baselineskip = 1.2\baselineskip
\global\parskip = 4pt plus 0.3pt
\global\abovedisplayskip = 18pt plus3pt minus9pt
\global\belowdisplayskip = 18pt plus3pt minus9pt
\global\abovedisplayshortskip = 6pt plus3pt
\global\belowdisplayshortskip = 6pt plus3pt


\def\endignore{}
\def\ignore #1\endignore{}

\newcount\dflag
\dflag = 0


\def\monthname{\ifcase\month
\or Jan \or Feb \or Mar \or Apr \or May \or June%
\or July \or Aug \or Sept \or Oct \or Nov \or Dec
\fi}

\def\timestring{{\count0 = \time%
\divide\count0 by 60%
\count2 = \count0
\count4 = \time%
\multiply\count0 by 60%
\advance\count4 by -\count0
\ifnum\count4 < 10 \toks1 = {0}
\else \toks1 = {} \fi%
\ifnum\count2 < 12 \toks0 = {a.m.}
\else \toks0 = {p.m.}
\advance\count2 by -12%
\fi%
\ifnum\count2 = 0 \count2 = 12 \fi
\number\count2 : \the\toks1 \number\count4%
\thinspace \the\toks0}}

\def\today{\ifcase\month\or January\or February\or March\or
 April\or May\or June\or July\or August\or September\or
 October\or November\or December\fi \space\number\day, \number\year}



\def\endtitle{}
\def\title#1\endtitle{\vskip.5in\titlefont
\global\baselineskip = 2\baselineskip
#1\vskip.4in
\baselineskip = 0.5\baselineskip\rm}
 
\def\endauthors{}
\def\authors#1\endauthors{#1}

\def\endabstract{}
\def\abstract#1\endabstract{\vskip .3in%
\centerline{\sectionfont\bf Abstract}%
\vskip .1in
\noindent#1}

\newcount\nsection
\newcount\nsubsection

\def\section#1{\global\advance\nsection by 1
\nsubsection=0
\bigskip\noindent\centerline{\sectionfont \bf \number\nsection.\ #1}
\bigskip\rm\nobreak}

\def\subsection#1{\global\advance\nsubsection by 1
\bigskip\noindent\sectionfont \sl \number\nsection.\number\nsubsection)\
#1\bigskip\rm\nobreak}

\def\topic#1{{\medskip\noindent $\bullet$ \it #1:}} 
\def\endtopic{\medskip}

\def\appendix#1#2{\bigskip\noindent%
\centerline{\sectionfont \bf Appendix #1.\ #2}
\bigskip\rm\nobreak}


\newcount\nref
\global\nref = 1

\def\ref#1#2{\xdef #1{[\number\nref]}
\ifnum\nref = 1\global\xdef\therefs{\noindent[\number\nref] #2\ }
\else
\global\xdef\oldrefs{\therefs}
\global\xdef\therefs{\oldrefs\vskip.1in\noindent[\number\nref] #2\ }%
\fi%
\global\advance\nref by 1
}

\def\listrefs{\vfill\eject\section{References}\therefs}


\newcount\nfoot
\global\nfoot = 1

\def\foot#1#2{\xdef #1{(\number\nfoot)}
\footnote{${}^{\number\nfoot}$}{\eightrm #2}
\global\advance\nfoot by 1
}


\newcount\nfig
\global\nfig = 1

\def\fig#1{\xdef #1{(\number\nfig)}
\global\advance\nfig by 1
}


\newcount\cflag
\newcount\nequation
\global\nequation = 1
\def\eqlabel{(1)}

\def\nexteqno{\ifnum\cflag = 0
\global\advance\nequation by 1
\fi
\global\cflag = 0
\xdef\eqlabel{(\number\nequation)}}

\def\lasteqno{\global\advance\nequation by -1
\xdef\eqlabel{(\number\nequation)}}

\def\label#1{\xdef #1{(\number\nequation)}
\ifnum\dflag = 1
{\escapechar = -1
\xdef\draftname{\littlefont\string#1}}
\fi}

\def\clabel#1#2{\xdef\eqlabel{(\number\nequation #2)}
\global\cflag = 1
\xdef #1{\eqlabel}
\ifnum\dflag = 1
{\escapechar = -1
\xdef\draftname{\string#1}}
\fi}

\def\cclabel#1#2{\xdef\eqlabel{#2)}
\global\cflag = 1
\xdef #1{\eqlabel}
\ifnum\dflag = 1
{\escapechar = -1
\xdef\draftname{\string#1}}
\fi}


\def\eeq{}

\def\eqnn #1\eeq{$$ #1 $$}

\def\eq #1\eeq{\xdef\draftname{\ }
$$ #1
\eqno{\eqlabel \rlap{\ \draftname}} $$
\nexteqno}



\def\eol{& \eqlabel \rlap{\ \draftname} \crcr
\nexteqno
\xdef\draftname{\ }}

\def\eeol{& \eqlabel \rlap{\ \draftname}
\nexteqno
\xdef\draftname{\ }}

\def\eolnn{\cr
\global\cflag = 0
\xdef\draftname{\ }}

\def\eeolnn{\xdef\draftname{\ }}

\def\eqa #1\eeq{\xdef\draftname{\ }
$$ \eqalignno{ #1 } $$
\global\cflag = 0}


\def\ie{{\it i.e.\/}}
\def\eg{{\it e.g.\/}}
\def\etc{{\it etc.\/}}
\def\etal{{\it et.al.\/}}


\def\anp#1#2#3{{\it Ann.~Phys. (NY)} {\bf #1} (19#2) #3}

\def\ijmp#1#2#3{{\it Int.~J.~Mod.~Phys.} {\bf A#1} (19#2) #3}

\def\mpla#1#2#3{{\it Mod.~Phys.~Lett.} {\bf A#1}, (19#2) #3}

\def\npb#1#2#3{{\it Nucl.~Phys.} {\bf B#1} (19#2) #3}
\def\plb#1#2#3{{\it Phys.~Lett.} {\bf #1B} (19#2) #3}
\def\pla#1#2#3{{\it Phys.~Lett.} {\bf #1A} (19#2) #3}

\def\prd#1#2#3{{\it Phys.~Rev.} {\bf D#1} (19#2) #3}
\def\pr#1#2#3{{\it Phys.~Rev.} {\bf #1} (19#2) #3}

\def\prl#1#2#3{{\it Phys.~Rev.~Lett.} {\bf #1} (19#2) #3}

\def\zpc#1#2#3{{\it Zeit.~Phys.} {\bf C#1} (19#2) #3}


\global\nulldelimiterspace = 0pt



\def\frac#1#2{{{#1} \over {#2}}\,}  
\def\hf{{1\over 2}}
\def\nth#1{{1\over #1}}


\def\Square{{\vbox {\hrule height 0.6pt\hbox{\vrule width 0.6pt\hskip 3pt
        \vbox{\vskip 6pt}\hskip 3pt \vrule width 0.6pt}\hrule height 0.6pt}}}
\def\Dsl{\hbox{/\kern-.6700em\it D}} 
\def\dsl{\hbox{/\kern-.5300em$\partial$}}
\def\pxpsl{\hbox{/\kern-.5600em$p$}}
\def\ssl{\hbox{/\kern-.5300em$s$}}
\def\epssl{\hbox{/\kern-.5100em$\epsilon$}}
\def\delsl{\hbox{/\kern-.6300em$\nabla$}}
\def\lxpsl{\hbox{/\kern-.4300em$l$}}
\def\elxpsl{\hbox{/\kern-.4500em$\ell$}}
\def\kxpsl{\hbox{/\kern-.5100em$k$}}
\def\qxpsl{\hbox{/\kern-.5000em$q$}}
\def\sla#1{\raise.15ex\hbox{$/$}\kern-.57em #1}
\def\Pl{\gamma_{\sss L}}



\def\twi{\widetilde}

\def\roughly#1{\mathrel{\raise.3ex\hbox{$#1$\kern-.75em\lower1ex\hbox{$\sim$}}}}
\def\lsim{\roughly<}
\def\gsim{\roughly>}

\def\tw#1{\tilde{#1}}
\def\ol#1{\overline{#1}}





\def\Scl{{\cal L}}






\def\hc{{\rm h.c.}}
\def\cc{{\rm c.c.}}

\def\sm{standard model}



\def\gem{$U_{\rm em}(1)$}
\def\em{{\rm em}}
\def\cc{{\rm cc}}
\def\nc{{\rm nc}}
\def\mw{M_{\sss W}}
\def\mz{M_{\sss Z}}
\def\gf{G_{\sss F}}

\def\rht{{\sss R}}
\def\lft{{\sss L}}

\def\twe{\tw{e}}

\def\sw{s_w}
\def\cw{c_w}

\def\dpi{\delta \Pi}

\def\dlambda{\delta \Lambda}
\def\ssa{\gamma}
\def\ssv{{\sss V}}
\def\ssw{{\sss W}}

\def\ssz{{\sss Z}}
\def\za{{\sss Z}\gamma}
\def\mw{M_{\sss W}}
\def\mz{M_{\sss Z}}
\def\smz{m_{\sss Z}}
\def\gf{G_{\sss F}}
\def\rht{{\sss R}}
\def\lft{{\sss L}}
\def\sm{{\sss SM}}
\def\tgv{{\sss TGV}}
\def\ww{{\sss WW}}
\def\aa{\gamma \gamma}
\def\zz{{\sss ZZ}}
\def\za{{\sss Z}\gamma}

\def\twe{\tw{e}}
\def\twmz{\tw{m}_\ssz}
\def\twmw{\tw{m}_\ssw}
\def\tws{\tw{s}_w}
\def\twc{\tw{c}_w}
\def\sw{s_w}
\def\cw{c_w}

\def\gz{g_\ssz}
\def\dgz{\Delta g_{1\ssz}}

\def\gv{g_\ssv}
\def\kv{\kappa_\ssv}
\def\dkv{\Delta \kv}
\def\dkz{\Delta \kappa_\ssz}
\def\dkg{\Delta \kappa_\gamma}
\def\lv{\lambda_\ssv}
\def\lz{\lambda_\ssz}
\def\lg{\lambda_\gamma}
\def\msbar{$\ol{\hbox{MS}}$}


\rightline{McGill-93/14}
\rightline{UdeM-LPN-TH-93-154}
\rightline{OCIP/C-93-7}
\rightline{hepph-9307223}
\rightline{June 1993}
\vskip .2in

\title
\centerline{BOUNDING ANOMALOUS}
\centerline{GAUGE-BOSON COUPLINGS}
\endtitle

\authors
\centerline{C.P.~Burgess,${}^a$ Stephen Godfrey,${}^b$ Heinz K\"onig,${}^b$
David London${}^c$ and Ivan Maksymyk${}^c$}  
\vskip .15in
\centerline{\it ${}^a$ Physics Department, McGill University}
\centerline{\it 3600 University St., Montr\'eal, Qu\'ebec, CANADA, H3A 2T8.}
\vskip .1in
\centerline{\it ${}^b$Ottawa-Carleton Institute for Physics}
\centerline{\it Physics Department, Carleton University}
\centerline{\it Ottawa, Ontario, CANADA, K1S 5B6.} 
\vskip .1in
\centerline{\it ${}^c$ Laboratoire de Physique Nucl\'eaire, l'Universit\'e de
Montr\'eal} 
\centerline{\it C.P. 6128, Montr\'eal, Qu\'ebec, CANADA, H3C 3J7.}
\endauthors

\abstract
We investigate indirect phenomenological bounds on anomalous
three-gauge-boson couplings. We do so by systematically determining their
one-loop implications for precision electroweak experiments. We find that
these loop-induced effects cannot be parametrized purely in terms of the
parameters $S$, $T$ and $U$. Like some other authors, we find many
cancellations among the loop-induced effects, and we show how to cast the
low-energy effective theory into a form which makes these cancellations
manifest at the outset. In a simultaneous fit of all CP conserving
anomalous three-gauge-boson couplings, our analysis finds only weak
phenomenological constraints. 
\endabstract


\vfill\eject
\section{Introduction}
The experimental couplings of the electroweak gauge bosons to light
fermions have now been quite well explored, particularly using low-energy
lepton scattering experiments and precision measurements at the $Z$
resonance. However, accurate experimental information is not available for
the self-interactions among the gauge bosons. This situation is likely to
be partially alleviated once the centre-of-mass energies at  the LEP
$e^+e^-$ collider are raised above the threshold for $W^\pm$ pair
production (LEP200). Given a sufficiently large sample of $W^\pm$ pairs,
direct information becomes available concerning the nature of the
$WW\gamma$ and $WWZ$ couplings. It is expected that a deviation (of 10\% or
more) of the three gauge-boson vertices (TGV's) from their standard model
(SM) values can be detected in this way
\ref\hagiwara{K. Hagiwara, R.D. Peccei, D. Zeppenfeld and K. Hikasa,
\npb{282}{87}{253}.}
\ref\detectable{D. Zeppenfeld, \plb{183}{87}{380}; 
D. Treille {\it et al.}, Proceedings of the ECFA Workshop on LEP200
vol. 2, p4.14, A. B\"ohm, W. Hoogland, (eds) Aachen (1986) CERN 87-08; 
D. Dicus and K. Kallianpur, \prd{32}{85}{32}.}
\ref\kane{G. Kane, J. Vidal, C.P. Yuan, \prd{39}{89}{2617}.}
\hagiwara, \detectable, \kane.

The key question is whether there are any kinds of new, nonstandard physics
that can give rise to this large a deviation from SM predictions for the
TGV's, and yet which might not have been detected elsewhere in other
low-energy experiments. As might have been expected, a great deal of effort
has been expended on researching this subject, leading to a dauntingly
large literature on the subject
\ref\oldbounds{A small sampling is:
F. Herzog, \plb{148}{84}{355};
M. Suzuki \plb{153}{85}{289};
J.C. Wallet, \prd{32}{85}{813};
A. Grau and J.A. Grifols, \plb{154}{85}{283}; \plb{166}{86}{233};
R. Alcorta, J.A. Grifols, and S. Peris, \mpla{2}{87}{23};
J.A. Grifols, S. Peris, and J. Sol\'a, \plb{197}{87}{437};
\ijmp{3}{88}{225};
P. Mery, S.E. Moubarik, M. Perrottet, and F.M. Renard, \zpc{46}{90}{229}; 
G. B\'elanger, F. Boudjema, and D. London, \prl{65}{90}{2943};
S. Godfrey and H. K\"onig, \prd{45}{92}{3196};
K.A. Peterson, \plb{282}{92}{207};
T. G. Rizzo, Argonne report ANL-HEP-PR-93-19 (1993; unpublished).}
\oldbounds. 

Two approaches to answering this question may be taken, depending on 
one's theoretical prejudices. 

\topic{Theoretically-Motivated Bounds}
The first approach is to use (sometimes fairly benign) assumptions about
the nature of the new physics in order to obtain an estimate of how large
the anomalous TGV's might be. The strength of this type of conclusion is
then inversely related to the restrictiveness of the assumptions that are
required in order to derive it. The thrust of this line of thinking is
usually to (reasonably convincingly) argue that induced anomalous TGV's are
unlikely to be larger than O(1\%) of their SM counterparts. If true, this
would make their detection at LEP improbable.

There are two broad classes of new physics for which this conclusion is
probably true. Firstly, if the new physics is perturbatively coupled to the
electroweak gauge bosons, then its contributions to TGV's are of order
$(g/4\pi)^2 \sim 10^{-3}$, where $g$ is an electroweak coupling constant.
Since the transverse gauge bosons couple with a universal strength, this
estimate includes a great many models. Any couplings between the new
physics and the longitudinal gauge bosons need not be so small, however,
and so a strongly-coupled symmetry-breaking sector might be considered.

In this case, a second line of reasoning leads to a similar conclusion
concerning the detectablity of anomalous TGV's 
\ref\einhorn{M. Einhorn \etal, preprints  UM-TH-93-12, UM-TH-92-25
(unpublished).}
\einhorn. To the extent that the
low-energy $W^\pm$ physics is dominated purely by the couplings of the
longitudinal modes, it may be parametrized using familiar techniques of
chiral perturbation theory
\ref\xpt{S. Coleman, J. Wess and B. Zumino, \pr{177}{69}{2239}; 
E.C. Callan, S. Coleman, J. Wess and B. Zumino, \pr{177}{69}{2247};  
J. Gasser and H. Leutwyler, \anp{158}{84}{142}.}
\xpt. It is therefore quite plausible that the size of these effective
interactions depends on the weak scale and the unknown scale, $M$, of the
new physics in the same way as do the corresponding couplings in the
low-energy chiral limit of QCD. This dependence may be succinctly
summarized by the rules of `Naive Dimensional Analysis' 
\ref\nda{A. Manohar and H. Georgi, \npb{234}{84}{189}; H.Georgi and
L.Randall, \npb{276}{86}{241}; H. Georgi, \plb{298}{93}{187}.}
\nda, 
which indicate that the relative size of anomalous and SM TGV's
should be $O(\mw^2/M^2)$. Again, for $M$ as large as a few TeV, as might be
expected for a strongly-interacting Higgs sector for example, we are led to
expect anomalous TGV's of $O(1\%)$. In either case, these are too small to
be observed. 

A complementary line of argument is based on naturalness 
\ref\derujula{A. de R\'ujula, M.B. Gavela, P. Hernandez and E. Mass\'o,
\npb{384}{92}{3}.}
\derujula. One way of phrasing this argument states that anomalous TGV's
should be of the same size as other new-physics contributions to the purely
gauge sector (e.g. `oblique' corrections) 
\ref\peskin{M.E. Peskin and T. Takeuchi, \prl{65}{90}{964}; 
D.C. Kennedy and P. Langacker, \prl{65}{90}{2967}.}
\ref\marciano{W.J. Marciano and J.L. Rosner, \prl{65}{90}{2963}}
\ref\peskinprd{M.E. Peskin and T. Takeuchi, \prd{46}{92}{381}.}
\peskin, \marciano, \peskinprd, since there are no symmetries that could
naturally enforce a relative size difference. Since these oblique
corrections are bounded by precision electroweak measurements to be $\lsim
O(1\%)$, so the argument goes, anomalous TGV's can also be expected to be,
at most, this large. 

\topic{Purely Phenomenological Bounds}
The second, more conservative tack that may be taken in determining the
potential size of anomalous TGV's is to put theoretical prejudices aside
and ask for purely phenomenological bounds. In this case we ask whether
anomalous couplings that are large enough to be detected can be already
excluded based on other low-energy measurements. The potential bounds of
this sort arise either due to the direct probing of nonstandard gauge-boson
couplings in hadron colliders 
\ref\hadrona{J. Alitti \etal, The UA2 Collaboration, \plb{277}{92}{194}.}
\ref\hadronb{J. Cortez, K. Hagiwara, F. Herzog, \npb{278}{86}{26};
M. Samuel \etal, \prl{67}{91}{9}; \prd{44}{91}{2064};
U. Baur and E.L. Berger, \prd{41}{90}{1476}.}
\ref\hadronc{U. Baur and D. Zeppenfeld, \npb{308}{88}{127};
D. Zeppenfeld and S. Willenbrock, \prd{37}{88}{1775};
U. Baur and D. Zeppenfeld, \plb{201}{88}{383}; \npb{308}{88}{127};
F. Pastore and M. Pepe, Proceedings of the Large Hadron Collider
Workshop, Aachen, Vol II, p. 106, 1990, CERN 90-10.}
\kane, \hadrona, \hadronb, \hadronc, or from their indirect influence on
precisely measured quantities, through loops. Although there is agreement
that existing hadron machines cannot rule out detectable nonstandard TGV's
at LEP200 \hadronb, the extraction of bounds from loops has been more
controversial
\ref\usesandabuses{C.P. Burgess and D. London, preprint McGill-92/05,
UdeM-LPN-TH-84 (to appear in {\it Phys. Rev.}{ \bf D}).} 
\usesandabuses. 

Here again the most recent analyses may be classified according to the
assumptions that are made.  Some have entailed a weakly-coupled framework
within which the standard-model gauge group is linearly realized at low
energies by including an explicit light physical Higgs particle
\ref\shortmad{K. Hagiwara, et al. MAD/PH 690 (1992).}
\ref\linearmad{K. Hagiwara, S. Ishihara, R. Szalapski and D. Zeppenfeld, 
preprint MAD/PH/737, UT-635, KEK-TH-356, KEK preprint 92-214 (unpublished)}
\derujula, \shortmad, \linearmad. In all of these studies it is found
that detectably large anomalous TGV's cannot be ruled out purely by
phenomenology. Other workers 
\ref\nonlinearcern{P. Hern\'andez and F.J. Vegas, \plb{307}{93}{116}.}
\ref\nonlin{D. Choudhury, P. Roy and R. Sinha, preprint TIFR-TH/93-08
(unpublished).} 
\nonlinearcern, \nonlin\ have instead not assumed the existence of a light
Higgs, in an effort to extract bounds that are less constrained by 
assumptions concerning how the gauge symmetry is realized. Again the
detectable TGV's are not ruled out on purely phenomenological grounds.

It is natural to ask why purely phenomenological bounds should be pursued
at all, given that reasonably persuasive theoretical arguments indicate
that detectable anomalous TGV's are unlikely. Our own point of view is that
neither a purely theoretical estimate, nor a purely phenomenological
analysis is sufficient in itself. We can only hope to learn anything if
{\it both} types of investigations are performed, since it is only through
the comparison of both, and their subsequent confrontation with the direct
measurements at LEP, that we learn something about the nature of any new
physics. 
\endtopic

The present paper is intended as a contribution to the purely
phenomenological line of thought. We wish here to determine the constraints
on CP-preserving anomalous TGV's, with an absolute minimum of assumptions
about the nature of the new physics from which they are generated. As with
the previous analyses of refs.~\derujula, \linearmad, \nonlinearcern\ and
\nonlin, we assume that the scale, $M$, that is associated with the new
physics is high enough in comparison with the weak scale, $\mw$, to justify
an effective-lagrangian treatment that is controlled by powers of $1/M$.
Integrating out the physics at scale $M$ generates a host of effective
interactions, including anomalous TGV's among others:
\eq
\Scl_{\rm eff}(\mu^2 = M^2) = \Scl_\sm + \Scl_\tgv + \Scl_{\rm rest}. \eeq
Here $\Scl_\sm$ is the SM lagrangian, $\Scl_\tgv$ represents the  anomalous
TGV interactions and $\Scl_{\rm rest}$ denotes all of the other effective
operators. Only some of the effective interactions in $\Scl_{\rm rest}$
contribute at tree level to well-measured low-energy observables, and so
only these are presently well-constrained by the data. All other operators
--- including TGV's in particular ---  do not contribute in this way and so
are only bounded to the extent that they generate the better-constrained
operators as the effective theory is run down from $\mu = M$ to the much
lower scales where the low-energy measurements are ultimately performed.
Our purpose here is to compute which operators are generated by TGV's in
this way, and so to indirectly bound their coefficients.

Before describing our conclusions, it is useful to orient our calculation
in relation to the others that have recently been performed. Our
calculation differs from those of refs.~\derujula, \shortmad\ and
\linearmad\ in that we assume that the dominant degrees of freedom that
govern the loop contribution of TGV's to low-energy observables are {\it
only} the presently known particles (including the top quark). In
particular, we do {\it not} assume a light Higgs boson and do {\it not}
linearly realize the electroweak gauge group. As was emphasized in
ref.~\usesandabuses, one can choose to realize this gauge symmetry
nonlinearly, or to ignore it completely (apart from its \gem-invariant
subgroup), and in both cases  one is led to precisely the same low-energy
effective lagrangian 
\ref\equivalence{J.M. Cornwall, D.N. Levin and G. Tiktopoulos,
\prd{10}{74}{1145};
M.S. Chanowitz, M. Golden and H. Georgi, \prd{36}{87}{1490}.}
\equivalence. The price to be paid is that the new-physics scale cannot be
arbitrarily large: $\mw/M \gsim g/4\pi$, or else perturbative unitarity is
lost.

Our analysis also differs in important ways from those of
refs.~\nonlinearcern\ and \nonlin. Perhaps the most basic difference lies
in the number of effective operators that are considered. We use the most
general effective interactions of ref.~\hagiwara\ that are consistent with
CP conservation. Our calculations therefore include the five effective
couplings, $\dkg$, $\dkz$, $\dgz$, $\lz$, and $\lg$ (see the following
section for detailed definitions). Our results may be compared to those of
ref.~\nonlin\ by taking $\dgz=0$, and to those of ref.~\nonlinearcern\ by
choosing $\lz = \lg = 0$. It should be pointed out that the neglect of
$\lz$ and $\lg$ in ref.~\nonlinearcern\ is what would be expected if the
new physics were to produce an effective theory that satisfies the 
power-counting rules \nda\ that have been found from experience with chiral
perturbation theory in QCD. We do not make this assumption here, however,
since it is not a generic feature of all underlying theories at scale
$M$.\foot\linexamp{For example, $\lz$ and $\dkz$ are the same size in a
linearly-realized effective theory.} 

A further difference with other workers arises because the authors of 
refs.~\derujula\ and \nonlinearcern\ also make explicit uses of quadratic
divergences in arriving at their bounds. In the present language, their 
calculation corresponds to an estimate of the size of direct contributions
of new physics to $\Scl_{\rm rest}$ at scale $\mu = M$, rather than of the
induced effects at low energies due to the TGV's defined at this scale
\usesandabuses. This distinction is less important in the linearly-realized
case, where the low-energy divergences are less severe.

A calculation of the purely low-energy loop effects of TGV's has recently
been made without the assumption of a light Higgs in ref.~\nonlin. These
authors use the formalism of oblique corrections, as parametrized by 
Peskin and Takeuchi's $S$, $T$ and $U$ \peskin, to obtain their low-energy 
bounds on TGV's. One of the points of the present paper, however, is that
such an analysis, based on $S$, $T$ and $U$ is not sufficiently general for
inferring the complete low-energy effects of anomalous TGV's. As we have
emphasized elsewhere
\ref\stuvwx{I. Maksymyk, C.P. Burgess and D. London, preprint McGill-93/13,
UdeM-LPN-TH-93-151, hepph-9306267 (unpublished).} 
\stuvwx, the Peskin-Takeuchi formalism applies only to the extent that
new-physics contributions to self-energies for gauge bosons can be
approximated as expansions in $q^2/M^2$, truncated at the linear term. The
calculations of ref.~\nonlin, as well as our own, indicate that this is not
what is obtained through loops from anomalous TGV's. Instead we obtain
self-energies of the form
\eq\label\formnonointro
\delta\Pi(q^2) = \omega \mw^2 + \alpha q^2 + {\beta q^4 \over \mw^2} + 
{\gamma q^6 \over \mw^4}
\eeq
with $\omega$, $\alpha$, $\beta$ and $\gamma$ all of the same order of
magnitude. The Peskin-Takeuchi parametrization is therefore insufficient,
and must be replaced by the more general formalism, recently derived in
ref.~\stuvwx, which involves three additional parameters, denoted $V$, $W$
and $X$. Our analysis here also differs from that of ref.~\nonlin\ in that
we compute not only the oblique corrections resulting from loop
integration, but also corrections to fermion-gauge-boson vertices. These
vertex corrections can contribute significantly to low-energy observables. 

Our procedure consists of computing how the anomalous TGV's appear in the
six observable parameters $S$-$X$ that, in practice, completely parametrize
all precision low-energy electroweak measurements. We can then perform a
fit, including all charged- and neutral-current data at low energy and at
the $Z$ peak. The strength of the conclusions we can draw from such a fit
depends crucially on our assumptions regarding which terms we include in
our effective lagrangian. If we follow the fairly common practice of
fitting for one anomalous TGV at a time, setting the others to zero, we
find that that some of the anomalous couplings can be constrained to be too
small to detect at LEP200. However, this is rather unrealistic -- real
models of underlying physics do not generate just one anomalous TGV at a
time. A simultaneous fit for the general case in which expressions
involving all five anomalous TGV's are fitted to data finds that the
constraints are weak, in fact no stronger than those from direct
measurements by UA2 \hadrona. 

It should be noted that even this is not the least restrictive assumption
one can make. Besides the contributions to the parameters $S$ through $X$
that are generated by low-energy loops involving the anomalous TGV's
defined at the scale of new physics, $\mu = M$, there are also typically
direct contributions to $S$-$X$ that are generated from $\Scl_{\rm rest}$.
Obviously, if cancellations are allowed among these two types of
new-physics contributions, any constraints on anomalous TGV's are lost. In
this sense, current precision electroweak data can never rule out the
possibility of anomalous TGV's large enough to be detected at LEP200.
Nevertheless, it is a useful exercise to fit for one anomalous TGV at a
time, since it does indicate the degree to which the discovery of anomalous
TGV's at LEP200 would require cancellations among the new-physics effects
for some low-energy observables. 

The paper is organized in the following way: In Section 2, we define what
we mean by an anomalous TGV lagrangian sector. We imagine that the entire
effective lagrangian, defined just below the threshold for new physics,
$\mu = M$, consists of the standard model lagrangian plus a supplementary
TGV sector. In Section 3, we begin to estimate the indirect effects of this
anomalous TGV sector on electroweak observables, by integrating out the
physics between  $\mu = M$, and the electroweak scale, $\mu \simeq 100$
GeV. We do so by using the renormalization group (RG) -- in the \msbar\
renormalization scheme -- to run our effective lagrangian between these two
scales. We follow in particular how the effective TGV interactions at $\mu
= M$ induce other effective interactions at the weak scale, that are
detectable in precision electroweak measurements. We call the resulting
weak-scale theory the `grown' lagrangian, and present expressions for the
coefficients of the `grown' operators in terms of the anomalous TGV's.
Superficially, this `grown' lagrangian appears to be complicated,
consisting of some 25 differents types of terms.

In Section 4, however, we show that this awkwardness is illusory, since the
freedom to redefine fields can be used to greatly reduce the number of
terms and to conceptually simplify the calculation of observables.
Concretely, this amounts to the use of the SM equations of motion to
re-express the effective interactions
\ref\fieldredefinition{This observation goes back more than ten years to
early applications of effective lagrangians to gravity and string theory,
see \eg, M. Green, J. Schwarz and E. Witten, {\it Superstring Theory}
(Cambridge University Press, 1987). \hfil\break
More recent applications in the SM literature may be found in
C.P. Burgess and J.A. Robinson, in {\it BNL Summer Study on CP Violation},
S. Dawson and A. Soni editors, (World Scientific, Singapore, 1991);
H. Georgi, \npb{361}{91}{339}.}
\fieldredefinition. Exploiting the leeway offered by this technique, we can
cast the `grown' lagrangian into either of two particularly simple forms.
On the one hand, we may cast all of the induced interactions as oblique
vacuum-polarization corrections, with the proviso that these necessarily
involve higher-derivative interactions. These make a direct application of
the Peskin-Takeuchi $STU$ formalism invalid, requiring instead the more
general analysis in terms of the parameters $S$ through $X$ of
ref.~\stuvwx. An alternative reformulation, which we use in the present
paper, casts the `grown' lagrangian into a form to which the $STU$
formalism can be applied directly and which also includes a small number of
(higher-derivative) fermion-fermion-gauge-boson interactions that do not
contribute to low-energy observables. Both forms are equivalent, and the
freedom to transform freely from one to another demonstrates how many terms
in the original grown lagrangian ultimately cancel in physical observables. 
For completeness of presentation, we also give here a very brief summary of
the $STU$ parametrization of new physics, upon which we then build to
extend the existing formalism as required by the present analysis. We
derive formulae for electroweak observables in terms of the familiar
$S$,$T$,$U$ and the new parameters $V$,$W$,$X$.

In Section 5 we then present expressions for these six parameters in terms
of the anomalous TGV couplings. We explain how these expressions can be
used to place phenomenological bounds on the TGV couplings  and we conclude
that such bounds are weak. In Section 6 we present the results of fits to
these expressions. We first consider the limit in which each TGV is
separately turned on at $\mu = M$, with all other effective interactions
being zero. We find that, with this somewhat unphysical assumption, the
data place strong phenomenological bounds on the TGV couplings that are of
order of several percent. However, when we fit for the five anomalous TGV's
simultaneously, we find that the bounds become considerably weaker, of
${\cal O}(1)$. 

Finally, in Section 7 we discuss the results, commenting in particular on
some rather startling cancellations in our final expressions. These
cancellations decrease the sensitivity of some observables to certain
anomalous TGV's. 

\section{The Anomalous TGV's}

Let us start by considering the effective lagrangian, as defined at the
scale of new physics, $\mu = M$, after the lightest of the heavy new
particles has been integrated out. In general many effective interactions
appear in this lagrangian, but we wish to focus in this paper only on a few
of these\foot\direct{Note that the terms we ignore include direct
new-physics contributions to $S$, $T$ and $U$ \peskin,
as well as to gauge-boson-fermion-fermion and other vertices.}:
\label\effectivelagrangian
\eq  \Scl = \Scl_\sm(\twe_i) + \Scl_{\tgv}, \eeq
where $\Scl_\sm(\twe_i)$ is the standard model (SM) lagrangian, in which
the as-yet-undetected particles -- in particular, the Higgs boson -- have
been integrated out. We place tildes on the three electroweak parameters
($\twe_i \equiv \{\twe, \tws (\equiv\sin \tw{\theta}_\ssw), \twmz\}$) that
appear in this lagrangian as a reminder that their values have been shifted
from their `standard' values, which we denote without tildes: $e, \sw,
\smz$. These `standard' values are the ones that are obtained by fitting
radiatively corrected SM expressions for observables to precise data. 

$\Scl_{\tgv}$ contains the anomalous CP-conserving TGV's which are the
focus of this study. Following 
\ref\gaemers{K.J.F. Gaemers and G.J. Gounaris, \zpc{1}{79}{259}.}
refs.~\hagiwara\ and \gaemers\ we write:
\label\lthreeb
\eq \eqalign{
\Scl_{\tgv} &= i \gz \, \dgz \; \Bigl( W_{\alpha\beta} W^{*\beta} -
W^*_{\alpha\beta} W^{\beta} \Bigr) Z^\alpha \cr
& \qquad \qquad + i \sum_{\ssv = \ssz, \gamma} \Bigl[ \gv \, \dkv \; 
W^*_\alpha W_\beta V^{\alpha\beta} + \gv \, {\lv  \over \mw^2} \; 
W^*_{\rho\mu} {W^\mu}_\sigma V^{\sigma\rho} \Bigr]. \cr} \eeq
Here $g_\gamma = e$ and $\gz = e \cw/\sw$ denote the SM couplings.
Electromagnetic gauge invariance requires that $\Delta g_{1\gamma} = 0$. A
gauge field having two lorentz indices, such as $W_{\alpha\beta}$, denotes
the abelian curl of the corresponding gauge potential, defined using the
appropriate electromagnetic gauge-covariant derivative: $W_{\alpha\beta} =
D_\alpha W_\beta - D_\beta W_\alpha$. Here $D_\alpha W_\beta =
\partial_\alpha W_\beta + ie A_\alpha W_\beta$. Following the convention
established in the literature, we scale the $\lv$ term to $\mw^2$, despite
the fact that it is a dimension-six operator and should more correctly have
a factor $M^2$ in the denominator. Due to the appearance of the
\gem-covariant derivatives, $\Scl_{\tgv}$ also contains four- and
five-point gauge-boson vertices: 
\eq \Scl_\tgv = \Scl_{3{\sss B}} + \Scl_{4{\sss B}} + \Scl_{5{\sss B}}\eeq
where $\Scl_{3{\sss B}}$ is simply $\Scl_\tgv$ with the replacement $D_\mu
\to \partial_\mu$, and $\Scl_{4{\sss B}}$ is given by
\label\fourpoint
\eq  \Scl_{4{\sss B}} = e \gz \, \dgz \Bigl[ W^*_\alpha W_\beta ( Z^\alpha
A^\beta + Z^\beta A^\alpha) - 2 W^*_\alpha W^\alpha A_\beta Z^\beta \Bigr]
+ ( \lv \; \hbox{terms}). \eeq
The `$\lv$ terms' here consist of those four-point interactions that are
generated from the $\lv$ terms in $\Scl_\tgv$. We do not list them
explicitly because, for present purposes, all graphs which use these rules
turn out to vanish. As for $\Scl_{5{\sss B}}$, such terms contribute only
in two-loop diagrams and are ignored here.  

\section{Loop Integration and the Low-Energy Lagrangian}

We next calculate the loop effects of the anomalous TGV's of the previous
section. We do so by running the effective interactions of the lagrangian
defined at the new-physics scale, $\mu \simeq M$, down to the weak scale, 
$\mu \simeq \mz$, where we can extract their consequences for precision 
experiments. We compute here the coefficients of the effective interactions
at the weak scale that are generated through this renormalization group
(RG) mixing with the operators in $\Scl_\tgv$. 

\subsection{Vacuum Polarization and Vertex Corrections}

The required diagrams fall into two categories: contributions to the
gauge-boson vacuum polarization  --- \ie\ `oblique' --- corrections,  
\fig\vacpolfig\
as shown in Fig.~\vacpolfig, and fermion--gauge-boson vertex corrections,
as shown in \fig\vertexfig\
Fig.~\vertexfig. We define the running of our operators using the \msbar\ 
renormalization scheme.  Here we simply quote the results, presenting only
the  coefficient of $[2/(4-n) - \gamma_{\sss E}  + \ln (4\pi\mu^2)]$, where
$n$ is the dimension of spacetime.  

Evaluating the graphs of Fig.~\vacpolfig, we find the following TGV-induced
contributions to the gauge-boson vacuum polarization tensors. With the
definition
\label\defsofpols
\eq
\dpi^{\mu\nu}_{ab}(q) = \eta^{\mu\nu} \, \dpi_{ab}(q^2) + q^\mu q^\nu \quad
\hbox{terms},
\eeq
where $a,b = W,Z$ and $\gamma$ , the coefficient of 
$[2/(4-n) - \gamma_{\sss E} + \ln (4\pi\mu^2)]$ in $\dpi_{ab}(q^2)$ is:
\eq \eqalign{
\dpi_{\aa} (q^2) &= \alpha_\ssa q^2 + \beta_\ssa {q^4 \over \mz^2} 
+ \gamma_\ssa {q^6 \over \mz^4}, \cr
\dpi_{\za} (q^2) &= \alpha_{\za} q^2 + \beta_{\za} {q^4 \over \mz^2} 
+ \gamma_{\za} {q^6 \over \mz^4}, \cr
\dpi_\ww (q^2) &= \omega_\ssw \mw^2 + \alpha_\ssw q^2 + \beta_\ssw 
{q^4 \over \mw^2} + \gamma_\ssw {q^6 \over \mw^4}, \cr
\dpi_\zz (q^2) &= \omega_\ssz \mz^2 +\alpha_\ssz q^2 + \beta_\ssz {q^4
\over \mz^2}  + \gamma_\ssz {q^6 \over \mz^4}, \cr} \eeq
where $\alpha_\ssz, \beta_\ssa$ \etc\ are given as functions of the
couplings $\dgz$, $\dkv$ and $\lv$ in Table I. Anticipating that these
anomalous couplings are small, we drop all terms past linear order in the
expressions in the table. The vacuum polarization diagrams have also been
calculated by the authors of ref.~\nonlin, and our results are in agreement
with theirs for diagrams involving $\dkv$ and $\lv$. They did not calculate
the diagrams with $\dgz$. 

\midinsert
$$\vbox{\tabskip=0pt \offinterlineskip
\halign to \hsize{\strut#& #\tabskip 1em plus 2em minus .5em&
\hfil$#$\hfil &#& \hfil$#$\hfil &#\tabskip=0pt\cr
\noalign{\hrule}\noalign{\smallskip}\noalign{\hrule}\noalign{\medskip}
&& \hbox{Coefficient} && \hbox{One-Loop Result} \, \times
(\alpha/4\pi\sw^2) \log(\mu'^2/\mu^2)&\cr   
\noalign{\medskip}\noalign{\hrule}\noalign{\medskip}
&& \alpha_\ssa (\mu^2)&& \sw^2 (6\dkg - 12 \lg) &\cr
&& \beta_\ssa  (\mu^2)&& \sw^2(-{2 \over 3}\,\dkg + 2\lg)/\cw^2 &\cr
&& \gamma_\ssa  (\mu^2)&& -\nth{6} \,\sw^2 \dkg / \cw^4 &\cr
&& \alpha_{\za} (\mu^2) && \cw\sw[3\dkz + 3\dkg -6\lz -6\lg +4\dgz] &\cr
&& \beta_{\za}  (\mu^2)&& (\sw/\cw)[- \nth{3}(\dkz +\dkg) + \lz +\lg -
{5\over 6}\dgz] &\cr && \gamma_{\za}  (\mu^2)&& - (\sw/12\cw^3)[\dkz +
\dkg] &\cr 
&& \omega_\ssw  (\mu^2)&& 3[\sw^2\dkg + (1+\cw^2 - 1/(2\cw^2))\dkz + (1+
\cw^2 - \hf \,\cw^4)\dgz] &\cr
&& \alpha_\ssw  (\mu^2)&& {5\over 3}\sw^2\dkg + \nth{3}(7+5\cw^2)\dkz
-6\sw^2\lg - 6(1+\cw^2)\lz +\hf(1+{40\over 3}\cw^2+ {17\over 3} \cw^4)\dgz
&\cr 
&& \beta_\ssw  (\mu^2)&& [-{5\over 6}(\sw^2 \dkg+ \cw^2\dkz) + 2(\sw^2\lg
+\cw^2 \lz) - {1\over 6}(2\cw^2+7\cw^4) \dgz]  &\cr
&& \gamma_\ssw  (\mu^2)&& - \nth{6} \,\cw^4 \dgz &\cr
&& \omega_\ssz  (\mu^2)&& 9 \cw^4 \, \dgz &\cr
&& \alpha_\ssz  (\mu^2)&& \cw^2[ 6 \dkz - 12 \lz +8 \dgz] &\cr
&& \beta_\ssz  (\mu^2)&& [-{2 \over 3}\dkz + 2\lz - {5 \over 3}\dgz] &\cr
&& \gamma_\ssz  (\mu^2)&& -(1/6\cw^2)\dkz &\cr
\noalign{\medskip}\noalign{\hrule}\noalign{\smallskip}\noalign{\hrule}
}}$$
\centerline{\bf TABLE I}
\medskip
\noindent {\eightrm One-loop results for the coefficients in the
gauge-boson vacuum polarization in terms of the various TGV couplings,
where the TGV couplings are defined at scale $\mu'$.}   
\endinsert

As for the vertex graphs of Fig.~\vertexfig, we obtain the following 
expressions for the $[2/(4-n) -\gamma_{\sss E} + \ln (4\pi\mu^2)]$
coefficient in the fermion-fermion-gauge-boson vertex corrections:
\label\corrs
\eq \eqalign{
\dlambda_\em (q^2) &= \left( p_\em^{(2)} \, {q^2 \over \mz^2} + p_\em^{(4)}
\, {q^4 \over \mz^4} \right) \; T_{3f} \; \Pl , \cr
\dlambda_\cc (q^2) &= \left(p_\cc^{(0)} +  p_\cc^{(2)} \,  {q^2 \over \mw^2} 
+ p_\cc^{(4)} \, {q^4 \over \mw^4}  \right) \; V_{ff'} \; \Pl, \cr
\dlambda_\nc (q^2) &= \left( p_\nc^{(2)} \, {q^2 \over \mz^2} + p_\nc^{(4)}
\, {q^4 \over \mz^4} \right) \; T_{3f} \; \Pl. \cr} \eeq
where we have normalized the vertex corrections such that standard model
tree-level vertices are corrected in the following manner:
\label\defsofcorrs
\eq \eqalign{
i\Lambda_\em^{\mu}(q^2) &= - ie \gamma^{\mu}
\left[ Q_f + {1\over \sw} \delta\Lambda_\em (q^2)\right], \cr
i\Lambda_\cc^{\mu}(q^2) &= -i{e\over \sqrt{2}\sw} \, \gamma^{\mu}
\left[ V_{ff'} \Pl + \delta\Lambda_\cc (q^2) \right], \cr
i\Lambda_\nc^{\mu}(q^2) &= -i{e\over \sw\cw} \, \gamma^{\mu}
\left[ T_{3f}\Pl - Q_{f}\sw^2 + \cw \delta\Lambda_\nc (q^2) \right]. \cr}
\eeq 
where $f,f'$ denote fermion type, $T_{3f}$ is the weak isospin, and $Q_f$
the electric charge. In the charged-current expression, $V_{ff'}$
represents the usual Cabibbo-Kobayashi-Maskawa (CKM) matrix in generation 
space when the external fermions are quarks, and is given by $\delta_{ff'}$
when they are leptons.

We give expressions for the $p^{(n)}$ coefficients as linear combinations
of the anomalous TGV couplings in Table II. Notice that since we have
calculated in the approximation that all fermions are massless, we have
`grown' only left-handed corrections to the standard model
fermion-fermion-gauge-boson couplings. 

\midinsert
$$\vbox{\tabskip=0pt \offinterlineskip
\halign to \hsize{\strut#& #\tabskip 1em plus 2em minus .5em&
\hfil$#$\hfil &#& \hfil$#$\hfil &#\tabskip=0pt\cr
\noalign{\hrule}\noalign{\smallskip}\noalign{\hrule}\noalign{\medskip}
&& \hbox{Coefficient} && \hbox{One-Loop Result} \, \times (\alpha/4\pi\sw^2) 
\log(\mu'^2/\mu^2)  &\cr  
\noalign{\medskip}\noalign{\hrule}\noalign{\medskip}
&& p_\nc^{(2)} (\mu^2) && (1/2\cw)[\dkz - 2\lz  + {5\over 3}\dgz] &\cr
&& p_\nc^{(4)}  (\mu^2)&& (1/12 \cw^3) \dkz & \cr
&& p_\cc^{(0)}  (\mu^2)&& {3\over 4}[(\cw^2-1)\dkz + \sw^2 \dkg + 
(\cw^2 - \cw^4)\dgz] &\cr
&& p_\cc^{(2)}  (\mu^2)&& {1\over 4}[{5 \over 3}(\cw^2 \dkz + \sw^2 \dkg)
-4(\cw^2 \lz + \sw^2 \lg) + (\cw^2 + {8\over 3}\cw^4) \dgz] &\cr
&& p_\cc^{(4)}  (\mu^2)&& \nth{12}\cw^4 \dgz & \cr
&& p_\em^{(2)} (\mu^2) && (\sw/2\cw^2)[\dkg - 2\lg] &\cr
&& p_\em^{(4)}  (\mu^2)&& (\sw/12 \cw^4) \dkg & \cr
\noalign{\medskip}\noalign{\hrule}\noalign{\smallskip}\noalign{\hrule}
}}$$
\centerline{\bf TABLE II}
\medskip
\noindent {\eightrm One-loop results for the coefficients in the
gauge-boson--fermion vertex corrections in terms of the various TGV
couplings.}
\endinsert

The above expressions are universal corrections because of our neglect of
all fermion masses. However, the one situation for which this assumption is
inadequate is the coupling of the down-type quarks to the photon and to the
$Z$ boson, since these involve virtual top quarks, whose mass is not
small.\foot\othersft{This necessity to keep track of the top-quark mass is
one of the differences between our calculation and that of ref.~\linearmad,
who find that the fermion masses do not affect the \msbar\ RG evolution at
the one-loop level in the linearly-realized effective theory.} Only the
$Zb\ol{b}$ vertex is of practical importance, though, because the only
process in which these interactions are  probed is in the decay of the $Z$
into $b\ol{b}$ pairs. For this observable the vertex  correction has the
form:
\label\bbzvertex
\eq
\dlambda_\nc^{b\ol{b}}\! (\!q^2\!) = \dlambda_\nc^{\rm univ.}\!( \! q^2 \! )
 + {1\over 64\pi^2}
|V_{tb}|^2
{e^2 \cw \over \sw^2} 
\left( {m_t^2 \over \mw^2} \right) 
\log \left({\mu'^2 \over \mu^2} \right) 
\left[ 3 \dgz + {1 \over 2 \cw^2}\dkz \left( {q^2 \over \mz^2} \right)
\right]  \Pl , \eeq
where $\dlambda_\nc^{\rm univ.}(q^2)$ is the result given in eq.~\corrs\ and
Table II. 

\subsection{The Weak-Scale Effective Lagrangian}

These expressions may be interpreted as contributions to the effective
lagrangian at lower energy scales. In the \msbar\ scheme the resulting
expressions for the induced couplings at scale $\mu$ may be obtained from
the table by simply multiplying the results of Tables I and II by
$y\equiv\ln((\mu' / \mu )^2)$, where the TGV couplings $\dkz$, $\dkg$ \etc\
are taken to be defined at scale $\mu'$.

We may therefore write out those terms that are `grown' in the low-energy
lagrangian at scale $\mu$, due to the appearance of $\Scl_\tgv$ at the
higher scale $\mu' = M$:
\label\grown
\eq  {1\over y}\;
\Scl_{\rm grown} = \Scl_{\rm vac.\,pol.} + \Scl_{\rm ver.\,corr.} +
\Scl_{\rm nonuniv.} 
\eeq
with $\Scl_{\rm nonuniv.}$ containing the non-universal $m_t$-dependent
contributions of eq.~\bbzvertex, while
\label\vacpollagrangian
\eq \eqalign{
\Scl_{\rm vac.\,pol.} &= {\omega_\ssz \mz^2 \over 2} \, Z_\mu Z^\mu +
{\alpha_\ssz \over 4} \, Z_{\mu\nu} Z^{\mu\nu} - {\beta_\ssz \over 4 \mz^2}
\, Z_{\mu\nu} \Square Z^{\mu\nu} + {\gamma_\ssz \over 4 \mz^4} \,
Z_{\mu\nu} \Square^2 Z^{\mu\nu} \cr
& \qquad + {\alpha_{\za} \over 2} \, Z_{\mu\nu} F^{\mu\nu} - {\beta_{\za}
\over 2 \mz^2} \, Z_{\mu\nu} \Square F^{\mu\nu} + {\gamma_{\za} \over 2
\mz^4} \, Z_{\mu\nu} \Square^2 F^{\mu\nu} \cr
& \qquad + {\alpha_\ssa \over 4} \, F_{\mu\nu} F^{\mu\nu} - {\beta_\ssa
\over 4 \mz^2} \, F_{\mu\nu} \Square F^{\mu\nu} + {\gamma_\ssa \over 4
\mz^4} \, F_{\mu\nu} \Square^2 F^{\mu\nu} \cr
& \qquad + \omega_\ssw \mw^2 \, W^*_\mu W^\mu +
{\alpha_\ssw \over 2} \, W^*_{\mu\nu} W^{\mu\nu} - {\beta_\ssw \over 2
\mw^2} \, W^*_{\mu\nu} \Square W^{\mu\nu} + {\gamma_\ssw \over 2 \mw^4} \,
W^*_{\mu\nu} \Square^2 W^{\mu\nu} \cr} \eeq
and
\label\vertexlagrangian
\eq \eqalign{
\Scl_{\rm ver.\,corr.} &= \Bigl( \sw \, j^\mu + \cw \, N^\mu \Bigr) 
\left( - p^{(2)}_\nc {\Square \over \mz^2} + p^{(4)}_\nc {\Square^2 \over
\mz^4} \right) Z_\mu \cr 
& \qquad + \Bigl( \sw \, j^\mu + \cw \, N^\mu \Bigr) 
\left( - p^{(2)}_\em {\Square \over \mz^2} + p^{(4)}_\em {\Square^2 \over
\mz^4} \right) A_\mu \cr 
& \qquad +  \left[ J^\mu \left( p^{(0)}_\cc - p^{(2)}_\cc {\Square \over
\mw^2} + p^{(4)}_\cc {\Square^2 \over \mw^4} \right) W^*_\mu + \hc \right].
\cr} \eeq 
Here $j^\mu$, $N^\mu$ and $J^\mu$ are respectively the total SM
electromagnetic, neutral and charged currents: 
\eqa\label\currentdefs
j^{\mu}&=-e\sum_f\bar{\Psi}_f\gamma^{\mu}Q_f\Psi_f ,\eol
J^{\mu}&=-{e \over \sqrt{2} \sw }
\sum_{ff'}\bar{\Psi}_f\gamma^{\mu}V_{ff'} \Pl \Psi_{f'} , \eol
N^{\mu}&=-{e \over \sw\cw} \sum_f\bar{\Psi}_f\gamma^{\mu}
\left[ T_{3f} \Pl  - Q_f \sw^2 \right] \Psi_f, \eeol\eeq
so that $\Scl_\sm = j^\mu A_\mu + N^\mu Z_\mu + ( J^\mu W^*_\mu + \hc) + ...$ 

To the extent that the loop effects of anomalous TGV's are universal in
their coupling to fermion generations, we see that it is possible to
express the `grown' fermion-gauge couplings in terms of linear combinations
of total SM currents. This is important, since it in turn allows -- through
the use of equations of motion -- a significant simplification of the
effective lagrangian. The exception to this universal form is in the
$m_t$-dependent interactions of $\Scl_{\rm nonuniv.}$, in which the $Z$
boson and the photon do not couple with the same strength to all
generations of down quarks. This particular case must be treated
separately, but does not affect the arguments of the subsequent sections.

\section{Simplifying the Effective Theory}

At this point one might be daunted by the fact that there are no fewer than
29 types of terms in this `grown' lagrangian, since this potentially makes
any further analysis very cumbersome. This complicated form for $\Scl_{\rm
grown}$ is illusory, however, since not all of the effective interactions
displayed in Eqs.~\vacpollagrangian\ and \vertexlagrangian\ are independent
of one another. The point here is that, due to the freedom to perform field
redefinitions, there are a great many ways of writing any particular term
in an effective lagrangian. In particular, any two operators which can be
transformed into one another using the SM equations of motion may also be
transformed into one another by a field redefinition \fieldredefinition. 

We may therefore use the SM equations of motion to simplify $\Scl_{\rm
grown}$, with two goals in mind. We wish to firstly reduce the number of
types of terms that we must consider. Next, in choosing which of the
remaining terms to keep as a basis of independent interactions, we wish to
take advantage, as much as possible,  of existing formalism and bounds on
various types of new physics that have been obtained by previous workers.
We therefore try, in particular, to cast the grown lagrangian into a form
that is amenable to an application of the $STU$ formalism of Peskin and
Takeuchi \peskin. As we shall see, this is not completely possible, due to
the appearance in  $\Scl_{\rm grown}$ of higher-derivative interactions,
and so a naive application of this formalism turns out to be incorrect.
Instead, along the lines of \stuvwx, we use an  extension of this formalism
that permits the treatment of more general $q^2$ dependence. 

\subsection{STU Formalism}

Before casting the `grown' lagrangian in a form amenable to an application
of the $STU$ formalism, let us rapidly review the $STU$ basics, following a
method that we have developed in 
\ref\bigfit{C.P. Burgess, S. Godfrey, H. K\"onig, D. London and I. Maksymyk,
preprint McGill-93/12, UdeM-LPN-TH-93-155, OCIP/C-93-6 (unpublished).}
ref.~\bigfit.  

Imagine supplementing the standard model by an `$STU$-sector', \ie\ by the
lowest-dimension effective two-point interactions for gauge bosons:
\label\stusector
\eqa 
 \Scl_{eff} & = \Scl_{\sss SM}(\tw{e}_i) + \Scl_{\sss STU}, \eolnn
\hbox{with}\qquad 
    \Scl_{\sss STU} &= - {A \over 4} \; F_{\mu\nu} F^{\mu\nu}
   - {B \over 2} \; W_{\mu\nu}^* W^{\mu\nu} - {C \over 4} \;
   Z_{\mu\nu} Z^{\mu\nu} + {G \over 2} \; F_{\mu\nu} 
   Z^{\mu\nu} \eolnn
   & \qquad  + w \, \twmw^2 \; W^*_\mu W^\mu + {z \over 2}
   \, \twmz^2 \; Z_\mu Z^\mu . \eeol \eeq
where $\twe_i$ are defined as in Eq.~\effectivelagrangian. This
parametrization implies that self-energies generated by new physics are of
the form
\label\presupposedform
\eq
\delta\Pi_a (q^2) = \omega_a + \alpha_a q^2,
\eeq
which is consistent with the idea that $q^2 \ll M^2$ (where $M$ is the
scale of new physics) and that $\delta\Pi_a(q^2)$ can be expanded as a
Taylor series in $q^2/M^2 $, truncated at the linear term. 

It is a simple matter to canonically normalize and diagonalize the
gauge-boson kinetic terms. Working to linear order in the small 
coefficients $A,B,\dots $, the required field redefinitions are 
\label\rescaling
\eqa A_\mu &\rightarrow\left( 1 - {A\over 2} \right) \; A_\mu + G \;
Z_\mu,\eol 
W_\mu &\rightarrow\left( 1 - {B\over 2} \right) \; W_\mu , \eol
Z_\mu &\rightarrow\left( 1 - {C\over 2} \right) \; Z_\mu . \eeol 
\eeq
The field transformations affect the electromagnetic, charged and neutral
current couplings and the mass terms, which now become:
\label\zerothem
\eqa  \Scl_{\rm em} &=  -\twe \left( 1 - {A\over 2} \right) \sum_f 
\ol{\Psi}_f \gamma^\mu Q_f \Psi_f  \; A_\mu , \eol 
\label\zerothcc
\Scl_{\rm cc} &=  -{\twe  \over \tws \sqrt{2}} \left( 1 - {B\over 2}
\right) \sum_{ff'} \twi{V}_{ff'} \;  \ol{\Psi}_f \gamma^\mu \Pl \Psi_{f'}
\; W^*_\mu + h.c., \eol  
\label\zerothnc
\Scl_{\rm nc} &= -{\twe \over \tws \twc} \left( 1 - {C\over 2} \right)
\sum_f \ol{\Psi}_f \gamma^\mu \left[ T_{3f} \Pl - Q_f  \tws^2  + Q_f \tws
\twc \, G \right] \Psi_{f'} \; Z_\mu , \eol
\label\bosonmasses
\Scl_{\sss W,Z} & = (1 + w - B) \, \twmw^2 \; W^*_\mu W^\mu +
\hf \, (1 + z - C)  \, \twmz^2 \; Z_\mu Z^\mu . \eeol 
\eeq
Our next step is to eliminate $\twe_i$ in favour of the `standard'
parameters $e_i$. To this end, we express our model's predictions for the
three precisely measured observables $\alpha$, $G_{F \sss}$ and $M_{\sss
Z}$ and equate these expressions to the standard model formulae:
\label\inputting
\eqa  4\pi\alpha & = 4\pi\alpha_{{\sss SM}}(\twe,\tws,\twmz)
 (1 - A) \equiv 4\pi\alpha_{{\sss SM}}(e,\sw,m_{\sss Z}),\eolnn
\gf &= G_{\sss F \; SM}(\twe,\tws,\twmz) (1 - w)
 \equiv G_{\sss F \; SM}(e,\sw,m_{\sss Z}),\eol     
M^2_{{\sss Z}}&= M^2_{{\sss Z \; SM}}(\twe,\tws,\twmz)(1 + z - C) 
\equiv M^2_{{\sss Z \; SM}}(e,\sw,m_{\sss Z}) . \eeolnn \eeq
Inverting these equations, we obtain
\eqa \label\inversion
\twe & = e \left( 1 + {A \over 2} \right), \eolnn
\tilde{m}^2_{\sss Z} & = m^2_{\sss Z}(1 - z + C), \eol
\tws^2 & = \sw^2 \left(1-{\cw^2\over\cw^2-\sw^2}(-A-z+C+w)\right).\eeolnn 
\eeq

With the substitution of  Eqs.~\inversion\ into the various lagrangian 
terms given in Eqs.~\zerothem, \zerothcc\ and \zerothnc, one obtains
interaction terms consisting of the usual standard model piece corrected by
a linear combination of the parameters $A$ through $z$. It turns out,
though, that these parameters can only appear in three linearly independent
combinations, a convenient choice being 
\eqa\label\STUdefined
\alpha S &= 4\sw^2\cw^2 (A - C) - 4\sw\cw (\cw^2 - \sw^2) G ,\eolnn
\alpha T &= w - z, \eol 
\alpha U &= 4\sw^2 (\sw^2 A - B + \cw^2 C - 2\sw\cw G), \eeolnn
\eeq
so that all the corrections can be expressed in terms of $S$, $T$ and $U$. 
After this algorithm has been performed, the interaction and $W$ boson
mass terms become:
\label\finalSTUformscc
\eqa\Scl_{\rm cc} &=  
     -{e \over \sqrt{2} \sw} \left( 1 - { \alpha S \over 4 ( \cw^2 -
  \sw^2)} + { \cw^2\; \alpha T \over 2 (\cw^2 - \sw^2)}  + {\alpha U \over
8 \sw^2} \right) \sum_{ff'} V_{ff'}\ol{\Psi}_f \gamma^\mu \Pl \Psi_{f'} \;
   W^*_\mu + h.c.,  \eol 
\label\finalSTUformsnc
\Scl_{\rm nc} &= 
   -{ e \over \sw \cw} \left(\!\! 1 + {\alpha T \over 2 }\!\right) \sum_f
   \ol{\Psi}_f \gamma^\mu \!\left[ T_{3f} \Pl - Q_f  
\left( \! \sw^2 + {\alpha S \over
   4 ( \cw^2 -  \sw^2)} - { \cw^2\sw^2 \; \alpha T \over \cw^2 - \sw^2}  
    \!\right) \right] \!\! \Psi_f  Z_\mu , \eol 
\label\finalSTUformswmass
\Scl_{\sss W} & = m_{\sss Z}^2 \cw^2 \left[ 1 - {\alpha S \over 2 ( \cw^2 -
   \sw^2)}  + {\cw^2 \; \alpha T \over  \cw^2 - \sw^2}+{\alpha U \over 4
   \sw^2}\right]  \; W^*_{\mu} W^{\mu}. \eeol 
\eeq

{}From Eqs.~\finalSTUformscc, \finalSTUformsnc\ and \finalSTUformswmass, it
is a simple matter to compute the $STU$ dependence in expressions for
various electroweak observables. Such expressions can then be compared to
experimental data in order to place constraints on $S$, $T$ and $U$. These
constraints can be translated into bounds on the parameters of the new
physics in terms of which $S$, $T$ and $U$ are computed. 

However, this formalism cannot be directly applied in the present TGV
analysis: as mentioned earlier, it is based on the assumption that
$\delta\Pi(q^2)$ can be expressed as a Taylor series in $q^2/M^2$ truncated
at the linear term, whereas in the present anomalous TGV application (with
non-linearly  realized gauge-symmetry) the new physics self-energies are of
the form 
\eq\label\formnono
\delta\Pi_{\sss W}(q^2) = \omega \mw^2 + \alpha q^2 + {\beta q^4 \over
\mw^2} + {\gamma q^6 \over \mw^4}
\eeq
with $\omega$, $\alpha$, $\beta$ and $\gamma$ all of the same order. 
Clearly the $STU$ formalism as it stands cannot be applied directly, a
point which is easy to miss if the definitions given in ref.~\marciano\ are
used: 
\eqa
\label\illusorydefs
\alpha S & = {4 \over \mz^2 }
\Bigl[ 
\cw^2\sw^2\left( \Pi_\zz (\mz^2) - \Pi_\zz (0) \right) \eolnn
& - \cw^2\sw^2\Pi_{\aa} (\mz^2) + \sw\cw (\sw^2 - \cw^2 )\Pi_{\za} (\mz^2) 
\Bigr] ,\eolnn
\alpha T &= {\Pi_{\sss WW}(0)\over\mw^2} - {\Pi_{\sss ZZ}(0)\over\mz^2},
\eol
\alpha U &= 4 \sw^2 \Bigl[
{1 \over \mw^2}\left( \Pi_\ww (\mw^2) - \Pi_\ww (0) \right)  \eolnn
& - {1\over\mz^2} \left( \sw^2 \Pi_{\aa} (\mz^2) + 
\cw^2 \left( \Pi_\zz (\mz^2) - \Pi_\zz (0) \right) +
2 \cw\sw\Pi_{\za} (\mz^2) \right) \Bigr].
\eeolnn\eeq
The point is that although these definitions make sense for arbitrary
vacuum polarizations, it turns out that it is impossible to express all
well-measured observables using just these three parameters, {\it unless}
the self-energy expansions are truncated at linear order in $q^2$. A method
for dealing with the case of general $q^2$ dependence in the vacuum
polarization is given in ref.~\stuvwx, where it is shown that, in practice,
the contributions to current precision electroweak measurements may be 
encapsulated into only six variables, called $(STUVWX)$. 

We now turn to the task of casting our effective lagrangian in the form
most amenable to an application of the methods presented above.
             
\subsection{Using Equations of Motion to Simplify the Effective Lagrangian}

As was mentioned earlier, the analysis presented in Section 3 seems
somewhat cluttered, since $\Scl_{grown}$ entails 29 types of terms,
including oblique corrections (dimension-six and -eight terms) as well as
vertex corrections. Moreover, the presence of higher-derivative operators
appears to hinder the direct application of the formalism presented in the
previous subsection. So, to overcome these difficulties, we exploit the
freedom to simplify these interactions using  the Euler-Lagrange equations
of motion derived from $\Scl_\sm(\twe_i)$, 
\eqa \label\standardmodelequationsofmotion
\partial_\mu (\partial^\mu A^\nu -\partial^\nu 
A^\mu)=&- \tilde{j}^\nu , \eolnn 
\partial_\mu (\partial^\mu Z^\nu -\partial^\nu 
Z^\mu)=&-\tilde{N}^\nu -\twmz^2 Z^\nu ,\eol
\partial_\mu (\partial^\mu W^\nu -\partial^\nu 
W^\mu)=&-\tilde{J}^\nu -\twmw^2 W^\nu. \eeolnn \eeq 
Moreover, one can also make liberal use of the freedom to integrate by
parts to relate otherwise disparate operators. 

It is useful to note that a certain leeway exists in transforming operators
with the above-mentioned methods.  For example, a general re-expression of
the operator $Z_{\mu\nu}\Square Z^{\mu\nu}$ would be 
\eqa \label\leeway
Z_{\mu\nu}\Square Z^{\mu\nu}\rightarrow &
-\eta\rho\mz^2 Z_{\mu\nu}Z^{\mu\nu} 
- 2\eta(1-\rho)\mz^4 Z_\mu Z^\mu 
- 2\eta(2 - \rho - \xi)\mz^2 Z_\mu N^\mu  \eolnn 
& + 2\eta\xi N_\mu\Square Z^\mu 
- 2\eta(1 - \xi)N_\mu N^\mu 
+ (1 - \eta) Z_{\mu\nu}\Square Z^{\mu\nu}, \eeol \eeq
where $\eta$ , $\rho$ and $\xi$ may take completely arbitrary values. We
choose these parameters to streamline the present analysis. In particular,
it is noteworthy that some of these parameters may be chosen to ensure the
absence of four-fermion operators, which would otherwise complicate the
calculation of observables. 

Even after removing the four-fermi interactions in this way, there still
remains a great deal of latitude in the form into which the effective
theory may be cast. For example, even though our present lagrangian
involves both vertex and self-energy corrections, we may use the equations
of motion to transform all new-physics effects purely into self-energy
corrections. In this case the techniques of ref.~\stuvwx\ may be used
directly to bound the new physics. In the remainder of this paper we choose
an alternative procedure. We instead recast our grown lagrangian into a
form consisting of an `$STU$'-sector (as in Eq.~\stusector), as well as a
small set of higher-derivative vertex  corrections (which, of course, do
not contribute to low-energy observables). The parametrization of
observables in terms of the six variables $STUVWX$  nevertheless emerges at
the end, as it must.
  
To obtain this final result requires specific choices for the operator
transformations such as Eq.~\leeway. We display the particular versions
that we use in Table III. It is to be noted that an operator in the right
column of this table is meant to be equivalent to the operator on the left
to within ($i$) total divergences, ($ii$) terms which vanish with the SM
equations of motion, and ($iii$) terms which do not contribute to the
well-measured physical processes we later consider. There are two varieties
of terms of this last type. Firstly, some interactions have Feynman rules
which are  proportional to $q^{\mu}q^{\nu}$ and therefore do not contribute
in the limit of massless fermions. Secondly, we omit terms such as
fermion-fermion-photon-photon interactions, which cannot play a role at
tree-level in the electroweak observables under consideration. 

\midinsert
$$\vbox{\tabskip=0pt \offinterlineskip
\halign to \hsize{\strut#& #\tabskip 1em plus 2em minus .5em&
\hfil$#$\hfil &#& \hfil$#$\hfil &#\tabskip=0pt\cr
\noalign{\hrule}\noalign{\smallskip}\noalign{\hrule}\noalign{\medskip}
&& \hbox{Original Operator} && \hbox{Transformed Version of Operator} &\cr   
\noalign{\medskip}\noalign{\hrule}\noalign{\medskip}
&& W^{*}_{\mu\nu}\Square^2 W^{\mu\nu}  &&
\mw^4 W^{*}_{\mu\nu}W^{\mu\nu} + (J_\mu\Square^2 W^{*\mu} + h.c.)
- \mw^2 ( J_\mu\Square W^{*\mu} + h.c. )&\cr
%
&&W^{*}_{\mu\nu}\Square W^{\mu\nu}  &&
\mw^2 W^{*}_{\mu\nu}W^{\mu\nu} + (J_\mu\Square W^{*\mu} + h.c.) &\cr
%
%
&&W^{*}_{\mu}J^{\mu} + W_{\mu}J^{*\mu}  &&
W^{*}_{\mu\nu}W^{\mu\nu}  - 2 \mw^2 W^{*}_\mu W^\mu &\cr
%
&&F_{\mu\nu}\Square^2 F^{\mu\nu} && 
2 j_\mu \Square^2 A^\mu &\cr
%
&&F_{\mu\nu}\Square F^{\mu\nu} &&
2 j_\mu \Square A^\mu &\cr
&&Z_{\mu\nu}\Square^2 F^{\mu\nu} &&
2 j_\mu \Square^2 Z^\mu &\cr
&&Z_{\mu\nu}\Square F^{\mu\nu} &&
2 j_\mu \Square Z^\mu &\cr
&&Z_{\mu\nu}\Square^2 Z^{\mu\nu} &&
\mz^4 Z_{\mu\nu}Z^{\mu\nu} + 2 N_\mu\Square^2 Z^{\mu} 
- 2 \mz^2  N_\mu\Square Z^{\mu} &\cr
&&Z_{\mu\nu}\Square Z^{\mu\nu} && 
-\mz^2 Z_{\mu\nu}Z^{\mu\nu}  
+ 2 \mz^2  N_\mu\Square Z^{\mu} &\cr
%
&&N_\mu\Square A^\mu  &&
j_\mu \Square Z^\mu + (1/2) \mz^2  Z_{\mu\nu}F^{\mu\nu} &\cr
&&N_\mu\Square^2 A^\mu  &&
j_\mu \Square^2 Z^\mu + \mz^2 j_\mu\Square Z^\mu &\cr
\noalign{\medskip}\noalign{\hrule}\noalign{\smallskip}\noalign{\hrule}
}}$$
\centerline{\bf TABLE III}
\medskip
\noindent {\eightrm Operator Transformations 
Used to Simplify the Grown Lagrangian.}   
\endinsert

\subsection{Final Form of the `Grown' Effective Lagrangian}

With the help of the transformations in Table III, $\Scl_{grown}$ 
can be recast as 
\eq \label\totalgrownprime
\Scl_{grown}^{\prime} = \Scl_{\sss STU}^{\prime} +\Scl_{\rm
ver.corr.}^{\prime} + \Scl_{\rm nonuniv.} 
\eeq 
where $\Scl_{\rm nonuniv.}$ contains the $m_t$-dependent effects, unchanged
from eq.~\grown, and
\eqa
\label\STUprime
{1\over y}\;\Scl_{\sss STU}^{\prime} =&\; - {A\over 4}F_{\mu\nu}F^{\mu\nu}
                        - {B\over 2}W^{*}_{\mu\nu}W^{\mu\nu}
                        - {C\over 4}Z_{\mu\nu}Z^{\mu\nu}\eolnn
                         &+{G\over 2}A_{\mu\nu}Z^{\mu\nu} 
                          + \mw^2 w W^{*}_{\mu}W^{\mu}
                          + {\mz^2 \over 2} z Z_{\mu}Z^{\mu}\eeol\eeq
while
\eqa\label\vercorrprime
{1\over y}\;\Scl_{\rm ver.corr.}^{\prime}=&\;
- {1 \over \mz^2} M N_{\mu}\Square Z^{\mu}
\, + \,{1 \over \mz^4} P N_{\mu}\Square^2 Z^{\mu} \eolnn
&- {1 \over \mz^2} {\sw\over\cw} (M - N) j_{\mu}\Square Z^{\mu}
\, + \,{1\over\mz^4}{\sw\over\cw}(P - Q) j_{\mu}\Square^2 Z^{\mu} \eolnn
%
%
&- {1 \over \mw^2} H (J_{\mu}\Square W^{*}_{\mu} + h.c.)
\, + \, {1 \over \mw^4} KJ_{\mu}\Square^2 W^{*}_{\mu} \eolnn
%
%
&- {1 \over \mz^2} D j_{\mu}\Square A^{\mu}
\, + \,{1 \over \mz^4} E N_{\mu}\Square^2 A^{\mu}
\eeol\eeq
with $A$,$B$, \etc\ defined in Table IV.

\midinsert
$$\vbox{\tabskip=0pt \offinterlineskip
\halign to \hsize{\strut#& #\tabskip 1em plus 2em minus .5em&
\hfil$#$\hfil &#& \hfil$#$\hfil &#\tabskip=0pt\cr
\noalign{\hrule}\noalign{\smallskip}\noalign{\hrule}\noalign{\medskip}
&& \hbox{Parameter from Eqs.~\STUprime\ and 
\vercorrprime} && \hbox{Definition} &\cr   
\noalign{\medskip}\noalign{\hrule}\noalign{\medskip}
&&A&& -\alpha_\ssa &\cr
&&B&&-\alpha_\ssw - \beta_\ssw - \gamma_\ssw - 2 p^{(0)}_\cc &\cr
&&C&&-\alpha_\ssz - \beta_\ssz - \gamma_\ssz &\cr
&&G&&+\alpha_{\za} - \cw p^{(2)}_\em &\cr
&&w&&\omega_\ssw - 2 p^{(0)}_\cc &\cr
&&z&&\omega_\ssz &\cr
&&M && {\beta_\ssz \over 2} + {\gamma_\ssz \over 2} + \cw p^{(2)}_\nc &\cr 
&&N && {\beta_\ssz \over 2}+{\gamma_\ssz \over 2}
-{\cw\over\sw} \left( \beta_{\za}  +  
\cw p^{(2)}_\em - \cw p^{(4)}_\em \right) &\cr 
&&P && {\gamma_\ssz \over 2} + \cw p^{(4)}_\nc &\cr
&&Q && {\gamma_\ssz \over 2} - 
{\cw\over\sw} \left( \gamma_{\za}
 + \cw p^{(4)}_\em \right)&\cr
%
%
&&H &&{\beta_\ssw \over 2} + {\gamma_\ssw \over 2} + p^{(2)}_\cc &\cr 
&&K &&{\gamma_\ssw \over 2} + p^{(4)}_\cc &\cr
%
%
&&D &&\sw p^{(2)}_\em + {\beta_{\ssa} \over 2} &\cr 
&&E &&\sw p^{(4)}_\em + {\gamma_{\ssa} \over 2} &\cr
\noalign{\medskip}\noalign{\hrule}\noalign{\smallskip}\noalign{\hrule}
}}$$
\centerline{\bf TABLE IV}
\medskip
\noindent {\eightrm Definitions of Parameters A,B,C, \etc\ Appearing in
Final Form of `Grown' Effective Lagrangian.}   
\endinsert

The final version to which we have converted the grown lagrangian contains
two categories of terms: mass and kinetic energy corrections for the gauge
bosons (i.e. an $STU$ sector) and a limited number of dimension-six and
dimension-eight vertex corrections. We have chosen this particular  form
for our grown lagrangian since the $STU$ formalism can now be applied
directly to the $STU$ sector in the calculation of expressions for 
observables. As for the vertex corrections in $\Scl_{grown}^{\prime}$, 
they do not contribute to low-energy observables, and it is an easy matter
to incorporate them in observables at other scales, especially since they
have been consolidated into a reduced number of terms. In particular, as
can be seen in Table III, we have decided to trade terms of type
$N^{\mu}\Square^n A_{\mu}$ for terms of type 
$j^{\mu}\Square^n Z_{\mu}$; we find that this
clarifies the calculation of neutral-current observables at the $Z$-pole by
allowing us to consider only Z-exchange diagrams.  

Importantly, this procedure hinges upon the fact that the grown currents
can be expressed as linear combinations of total standard model currents, a
property of all of the $TGV$-induced interactions except the
$m_t$-dependent terms of $\Scl_{\rm nonuniv.}$.

\vfill\eject
\section{Expressions for Observables}

In this section, we show how the six parameters ($S$, $T$, $U$, $V$, $W$
and $X$ of ref.~\stuvwx) emerge in the present context. We do so by
directly expressing the electroweak observables in terms of the
coefficients in our effective lagrangian.

Since the calculation of the input observables $\alpha$, $G_{\sss F}$ and
$\mz^2$ involves none of the higher-derivative vertex corrections,
Eqs.~\inputting\ and \inversion\ remain valid. Thus, using these results,
we may immediately write the final total form of the various interaction
sectors in terms of the canonically normalized fields. This entails simply
adding the dimension-six and -eight vertex corrections to the pieces
displayed in Eqs.~\finalSTUformscc\ and \finalSTUformsnc: 
\label\finalinttermsnc
\eqa
\Scl_\nc = &\;-{e\over \cw\sw}\Biggl[ \left( 1 + {\alpha T \over 2} \right) 
\ol{\Psi}_f \gamma_{\mu}\left[
T_{3f}\gamma_{\sss L} - Q_f
\left(\sw^2 + {\alpha \over 4 (\cw^2 - \sw^2)}S
       - {\cw^2\sw^2\alpha \over (\cw^2 -\sw^2)}
          T\right)\right]\,\Psi_f \,Z^{\mu} \eolnn 
- &\ol{\Psi}_f \gamma_{\mu}\left[T_{3f}M\gamma_{\sss L} - 
      Q_f \sw^2 N \right] \Psi_f\,{\Square \over \mz^2}Z^{\mu}
+ \ol{\Psi}_f \gamma_{\mu}\left[T_{3f}P\gamma_{\sss L} - 
      Q_f \sw^2 Q \right] \Psi_f\,{\Square^2 \over \mz^4}Z^{\mu}
\Biggr], \eol
\label\finalinttermscc
\Scl_\cc = &\; J_{\mu}\left(
1 - {\alpha \over 4 (\cw^2 - \sw^2)}S
       - {\sw^2\alpha \over 2 (\cw^2 -\sw^2)}T
       + {\alpha \over 8 \sw^2}U    
       - H\,{\Square\over \mw^2}
       + K\,{\Square^2 \over \mw^4}\right)W^{*\mu} + h.c., \eol
\label\finalinttermswmass
\Scl_{{\sss W}}=&\;
    \smz^2 \cw^2 \left[ 1 - {\alpha S \over 2 ( \cw^2 - \sw^2)}  + 
   {\cw^2 \; \alpha T \over  \cw^2 - \sw^2}+{\alpha U \over 4 \sw^2}\right] 
 \; W^*_{\mu} W^{\mu} , \eol
\label\finalinttermsem
\Scl_\em =&\; j_{\mu} \left(
1 - D\;{\Square\over\mz^2} + E\;{\Square^2\over\mz^4}\right)\;A^{\mu}.
\eeol\eeq
It is now a simple matter to use these interactions to derive expressions
for various electroweak observables:
\topic{Asymmetries}
{}From Eq.~\finalinttermsnc\ , we deduce the Feynman rule
\label\ncfeynmanrule\eqa
i\Lambda_{{\sss Zff}}^{\mu}(q^2)=&\;\;
-i{e \over \cw\sw}
\left(  1 + {\alpha T \over 2} \right) 
\gamma^{\mu}\eolnn
& \Biggl[ T_{3f}\left( 1 +  M\,{q^2 \over \mz^2} + P\,{q^4 \over
\mz^4}\right) 
\eolnn
& - Q_f
\left( \sw^2 + {\alpha \over 4 (\cw^2 - \sw^2)}S
       - {\cw^2\sw^2\alpha \over (\cw^2 -\sw^2)}T 
+\sw^2\left( N \, {q^2 \over \mz^2} + Q {q^4 \over \mz^4}\right) 
\right) \Biggr]. 
\eeol\eeq                                            
Forward-backward asymmetries and left-right asymmetries depend only on the 
`effective' $(\sw)_{eff}(q^2)$, which can be read off from
Eq.~\ncfeynmanrule: 
\eq
(\sw^2)_{eff}(q^2) = 
\sw^2 + {\alpha \over 4 (\cw^2 - \sw^2)}S
       - {\cw^2\sw^2\alpha \over (\cw^2 -\sw^2)}T  
+ \sw^2 
\left(
(N - M){q^2\over \mz^2} + (Q - P){q^4\over \mz^2}
\right).
\eeq
In particular, 
\label\sinatzpole
\eq
(\sw^2)_{eff}(\mz^2) = 
\sw^2 + {\alpha \over 4 (\cw^2 - \sw^2)}S
 - {\cw^2\sw^2\alpha \over (\cw^2 -\sw^2)}T  + \alpha X ,
\eeq
where we define the new parameter $X$ according to
\label\Xdefined
\eq
\alpha X  \equiv \sw^2 ( N + Q - M - P ).
\eeq
For the specific case of asymmetries in the decay $Z \to b \ol{b}$, the
effects of the nonuniversal $m_t$-dependent interactions must also be
computed. Since these are not universal, they cannot be parameterized
solely in terms of the variables $S$ through $X$. Their contribution to $Z
\to b \ol{b}$ asymmetries can be included by replacing $X$ by $\hat{X}$,
where
\label\hatxdef
\eq 
\hat{X} = X +  
{\cw^2 \over 8\pi} \,
|V_{tb}|^2 \,
\left( { m_t^2 \over \mw^2} \right)\,
\left[ 3 \dgz + { 1 \over 2\cw^2 } \dkz \right] 
\; \log\left( {\mu'^2\over \mu^2} \right).
 \eeq
\topic{Z-Decay Widths}
{}From Eq.~\ncfeynmanrule\ we obtain the partial width for $Z\rightarrow
\ol{f}f$: %
\label\zdecayformula
\eq
\Gamma_{Zf}=
\;{\mz\over 24\pi}\;{e^2\over \cw^2\sw^2}
( 1 + \alpha T + \alpha V )
\Bigl[ 
\Bigl( T_{3f} - Q_{f}(\sw^2)_{eff}(\mz^2) \Bigr)^2 +
\Bigl( Q_{f}(\sw^2)_{eff}(\mz^2)  \Bigr)^2 \Bigr],
\eeq
where we define the new parameter $V$ to be 
\label\Vdefined
\eq
\alpha V  \equiv 2 ( M + P ). \eeq
Again, the specific decay $Z \to b\ol{b}$ gets an additional contribution
due to the $m_t$-dependence of the $TGV$-induced lagrangian. It may be
incorporated by replacing $X$ by $\hat{X}$, as defined in eq.~\hatxdef, in
addition to making the replacement of $V$ by
\label\hatvdef
\eq
\hat{V} = V - 
{\cw^2 \over 4\pi\sw^2}\,
|V_{tb}|^2 \, 
\left( { m_t^2 \over \mw^2 }\,
\right) \left[ 3 \dgz + {1\over 2\cw^2} \dkz \right]  \; 
\log\left( {\mu'^2\over \mu^2} \right).
 \eeq
\topic{W-Decay Width}
{}From Eq.~\finalinttermscc\ one obtains the Feynman rule
\label\ccfeynmanrule
\eq
i\Lambda^{\mu}_{\nu l {\sss W}}(q^2) =  
-i {e \over \sqrt{2} \sw }\gamma^{\mu}
\left(
1 - {\alpha \over 4 (\cw^2 - \sw^2)}S
       - {\sw^2\alpha \over 2 (\cw^2 -\sw^2)}T
       + {\alpha \over 8 \sw^2}U    
       + H {q^2\over \mw^2} + K {q^4\over \mw^4} 
\right).
\eeq
Thus the partial width for $W\rightarrow l\nu_l$ is
given by 
\label\wdecayformula
\eq
\Gamma_{\ssw} = 
{\mw \over 24 \pi}{e^2 \over 2 \sw^2}
\left(
1 - {\alpha \over 2 (\cw^2 - \sw^2)}S
       - {\sw^2\alpha \over (\cw^2 -\sw^2)}T
       + {\alpha \over 4 \sw^2}U + \alpha W
\right) ,
\eeq
where we define the new parameter
\label\Wdefined
\eq
\alpha W \equiv 2 ( H + K ).
\eeq
\topic{W Mass}
The expression for W mass can be read off directly from
Eq.~\finalSTUformswmass:
\eq\label\wmass
M_{\sss W}^2 = \left( M_{\sss W}^2 \right)_{\sss SM} 
\left[ 1 - {\alpha S \over 2 ( \cw^2 - \sw^2)}  + 
   {\cw^2 \; \alpha T \over  \cw^2 - \sw^2}+{\alpha U \over 4 \sw^2}
\right]. 
\eeq
\topic{Low-Energy Observables}
Let us now turn to the low-energy experiments such as deep-inelastic
scattering and atomic parity-violation.  They are insensitive to the $q^2$
and $q^4$ pieces in the Eq.~\finalinttermsnc\ which originate from the 
dimension-six and -eight vertex corrections in $\Scl_{grown}^{\prime}$. 
Therefore, these observables involve only $S$, $T$ and $U$, and the
appropriate formulae can be taken directly from the usual $STU$ treatment
(see Appendix B in \peskinprd).

A comprehensive list of expressions for the electroweak observables that we
include in our analysis is given in Table V. These expressions consist of 
a radiatively corrected standard model prediction plus a linear combination 
of the six parameters $S$, $T$, $U$, $V$, $W$ and $X$. $\Gamma_Z$ and
$\Gamma_{b\bar{b}}$ are the total width of the $Z$ and its partial width
into $b\bar{b}$, respectively; $A_{FB}(f)$ is the forward-backward
asymmetry for $e^+e^- \to f\bar{f}$; $A_{pol}(\tau)$, or $P_\tau$, is the
polarization asymmetry  defined by $A_{pol}(\tau) = (\sigma_\rht -
\sigma_\lft)/ (\sigma_\rht + \sigma_\lft)$, where $\sigma_{\sss L,R}$ is
the cross section for a correspondingly polarized $\tau$ lepton;
$A_e(P_\tau)$ is the joint forward-backward/left-right asymmetry as
normalized in 
\ref\paul{P. Langacker, to appear in the 
{\sl Proceedings of 30 Years of Neutral Currents}, Santa Monica, 
February 1993.}
ref.~\paul; 
and $A_{LR}$ is the polarization asymmetry which has been measured 
by the SLD collaboration at SLC 
\ref\sld{K. Abe {\sl et al.}, \prl{70}{93}{2515}.}
\sld. The low energy  observables $g_L^2$ and $g_R^2$ are measured in deep
inelastic $\nu N$ scattering, $g^e_V$ and $g^e_A$ are measured in $\nu e
\to \nu e$ scattering, and $Q_W(Cs)$ is the weak charge measured in atomic
parity violation in cesium. The expressions for the low energy observables
are derived in refs.~\peskinprd\ and \bigfit.  

\midinsert
$$\vbox{\tabskip=0pt \offinterlineskip
\halign to \hsize{\strut#& #\tabskip 1em plus 2em minus .5em&#\hfil 
&#\tabskip=0pt\cr
\noalign{\hrule}\noalign{\smallskip}\noalign{\hrule}\noalign{\medskip}
&& \hfil \hbox{Expressions for Observables} &\cr  
\noalign{\medskip}\noalign{\hrule}\noalign{\medskip}
&& $\Gamma_\ssz  =(\Gamma_\ssz)_{\sss SM} - 0.00961 S + 0.0263 T
 + 0.0166 V - 0.0170 X + 0.00295 \hat{V} - 0.00369 \hat{X}$  (GeV)  & \cr
&& $\Gamma_{b\ol{b}}  =(\Gamma_{b\ol{b}})_{\sss SM} - 0.00171 S + 0.00416 T
 + 0.00295 \hat{V} - 0.00369 \hat{X}$ (GeV)  & \cr
&& $A_{\sss FB}(\mu)= (A_{\sss FB}(\mu))_{\sss SM} - 0.00677 S + 
0.00479 T - 0.0146 X$  & \cr
&& $A_{pol}(\tau) = (A_{pol}(\tau))_{\sss SM} -0.0284 S + 0.0201 T - 0.0613
X $  & \cr
&& $A_e (P_\tau) =(A_e(P_\tau))_{\sss SM} -0.0284 S + 0.0201 T - 0.0613 X $
&  \cr 
&& $A_{\sss FB}(b)=(A_{\sss FB}(b))_{\sss SM} -0.0188 S + 0.00984 T -0.0406
\hat{X}$ & \cr 
&& $A_{\sss FB}(c)=(A_{\sss FB}(c))_{\sss SM} -0.0147 S + 0.0104 T -0.03175
X$  &\cr 
&& $A_{\sss LR} =(A_{\sss LR})_{\sss SM} -0.0284 S + 0.0201 T - 0.0613 X  $
& \cr 
&& $M_\ssw^2=(M_\ssw^2)_{\sss SM}(1-0.00723 S +0.0111 T +0.00849 U)$ &\cr
&& $\Gamma_\ssw =(\Gamma_\ssw)_{\sss SM}(1-0.00723 S -0.00333 T + 0.00849 U
+ 0.00781W) $ & \cr
&& $g_{\sss L}^2 =(g_{\sss L}^2)_{\sss SM}-0.00269 S + 0.00663 T$ & \cr
&& $g_{\sss R}^2 =(g_{\sss R}^2)_{\sss SM} +0.000937 S - 0.000192  T$ & \cr
&& $g_{\sss V}^e(\nu e \to \nu e) =(g_{\sss V}^e)_{\sss SM} + 0.00723 S -
0.00541 T$  & \cr
&& $g_{\sss A}^e (\nu e \to \nu e) =(g_{\sss A}^e)_{\sss SM} - 0.00395 T$ &
\cr 
&& $Q_\ssw(^{133}_{55}Cs) = Q_\ssw(Cs)_{\sss SM} -0.795 S -0.0116 T$ & \cr
\noalign{\medskip}\noalign{\hrule}\noalign{\smallskip}\noalign{\hrule}
}}$$
\centerline{{\bf TABLE V}}
\medskip
\noindent {\eightrm Summary of the dependence of electroweak observables
on $S,T,U,V,W$ and $X$. In preparing this table we used the numerical values 
$\alpha(\mz^2)=1/128$ and $\sw^2=0.23$.}
\endinsert

There are several features in Table V worth pointing out. First, only  the
two parameters $S$ and $T$ contribute to the observables for which $q^2\sim
0$.  The parameter $U$ appears only in $\mw$ and $\Gamma_\ssw$. Given the
present uncertainty in $\Gamma_\ssw$, the limit on $U$ comes from the $M_W$
measurement. The parameter $W$ is weakly bounded, as it contributes only to
$\Gamma_\ssw$ which is at present poorly measured. In addition to $S$ and
$T$, observables on the $Z^0$ resonance are also sensitive to $V$ and $X$,
which are expressly defined at $q^2=\mz^2$. Observables that are not
explicitly given in Table V can be obtained using the given expressions. In
particular the parameter $R$ is defined as 
$R= \Gamma_{had}/\Gamma_{l\bar{l}}$, and $\sigma^h_p =
12\pi\Gamma_{e\bar{e}}\Gamma_{had}/\mz^2\Gamma_\ssz^2$ is the hadronic
cross section at the $Z$-pole.  

\vfill\eject
\section{Constraints on the Anomalous TGV Couplings}

We are now in a position to determine the phenomenological constraints on
the anomalous TGV's. To do so, we must first consider how the well-measured
couplings in the weak-scale effective lagrangian depend on the TGV
parameters defined at the high scale, $\mu' = M$. 

In the previous section, we have derived (Table V) expressions for
observables in terms of $S$ through $X$. To obtain formulae for $S$, $T$,
$U$, $V$, $W$ and $X$ in terms of the anomalous TGV's one substitutes the
expressions in Tables I and II into those of Table IV, and then uses the
definitions \STUdefined, \Xdefined, \Vdefined\ and \Wdefined. The result of
this substitution is displayed in Table VI, where we have taken $\mu' = M =
1$ TeV and $\mu = 100$ GeV. 

\midinsert
$$\vbox{\tabskip=0pt \offinterlineskip
\halign to \hsize{\strut#& #\tabskip 1em plus 2em minus .5em&
\hfil$#$\hfil &#& \hfil$#$\hfil &#\tabskip=0pt\cr
\noalign{\hrule}\noalign{\smallskip}\noalign{\hrule}\noalign{\medskip}
&& \hbox{Parameter} && \hbox{One-Loop Result} &\cr  
\noalign{\medskip}\noalign{\hrule}\noalign{\medskip}
&& S && 2.63 \dgz - 2.98 \dkg + 2.38 \dkz + 5.97 \lg - 4.50\lz &\cr
&& T && -1.82 \dgz + 0.550 \dkg + 5.83 \dkz &\cr
&& U && 2.42 \dgz - 0.908 \dkg - 1.91 \dkz + 2.04 \lg - 2.04 \lz &\cr
&& V && 0.183 \dkz  &\cr
&& W && 0.202 \dgz  &\cr
&& X && -0.0213 \dkg - 0.0611 \dkz &\cr
&& \hat{V}&& 0.183 \dkz - (3.68 \dgz + 0.797 \dkz) (m_t^2/\mw^2)  &\cr
&& \hat{X}&& -0.0213 \dkg - 0.0611 \dkz + 
(0.423 \dgz + 0.0916 \dkz) (m_t^2/\mw^2)   &\cr
\noalign{\medskip}\noalign{\hrule}\noalign{\smallskip}\noalign{\hrule}
}}$$
\centerline{\bf TABLE VI}
\medskip
\noindent {\eightrm One-loop results for the induced parameters $S$, $T$,
$U$, $V$, $W$ and $X$, defined at $\mu = 100$ GeV, in terms of the various
TGV couplings defined at $M=1$ TeV.  We have used $\alpha(\mz^2)=1/128$ and 
$\sw^2=0.23$.}
\endinsert

To obtain constraints on the anomalous TGV's, we perform a global fit. The
required expressions are obtained by substituting the results of Table VI
into those of Table V. The TGV-dependence of the nonuniversal,
$m_t$-dependent terms is given by eqs.~\hatxdef\ and \hatvdef. 

The experimental values and standard model predictions of the observables
used in our fit are given in Table VII. The standard model values have been
calculated with $m_t=150$ GeV and $M_H=300$ GeV. The LEP observables in
Table VII were chosen as they are closest to what is actually measured and
are relatively weakly correlated. In our analysis we include the combined
LEP values for the correlations 
\ref\correlations{The LEP Collaborations: ALEPH, DELPHI, L3, and OPAL,
\plb{276}{92}{247}.}
\correlations.
\ref\lep{C. DeClercqan, Proceedings of the Recontre de Moriond, Les 
Arcs France, March 1993; V. Innocente, {\it ibid}. }

\midinsert
$$\vbox{\tabskip=0pt \offinterlineskip
\halign to \hsize{\strut#& #\tabskip 1em plus 2em minus .5em&\hfil#\hfil 
&\hfil#\hfil &\hfil#\hfil &#\tabskip=0pt\cr
\noalign{\hrule}\noalign{\smallskip}\noalign{\hrule}\noalign{\medskip}
&& \hfil \hbox{Quantity} & \hfil \hbox{Experimental Value}& 
\hfil \hbox{Standard Model Prediction} &\cr  
\noalign{\medskip}\noalign{\hrule}\noalign{\medskip}
&& $M_Z$ (GeV)  & $91.187 \pm 0.007 $ \lep & input & \cr
&& $\Gamma_Z$ (GeV)  & $ 2.488 \pm 0.007 $ \lep & $2.490 [\pm0.006]$ & \cr
&& $R=\Gamma_{had}/\Gamma_{l{\bar l}}$  & $20.830 \pm 0.056$ \lep
	& $20.78 [\pm 0.07]$ & \cr
&& $\sigma^h_p $ (nb)  & $41.45 \pm 0.17$ \lep & $41.42 [\pm 0.06]$ & \cr
&& $\Gamma_{b\bar b}$ (MeV)  & $383 \pm 6$ \lep & $ 375.9 [\pm 1.3]$ & \cr
&& $A_{FB}(\mu)$ & $0.0165 \pm 0.0021$ \lep & $0.0141$ & \cr
&& $A_{pol}(\tau)$  & $0.142 \pm 0.017$ \lep & $ 0.137$ & \cr
&& $A_e (P_\tau)$  & $ 0.130 \pm 0.025$ \lep & $ 0.137$ & \cr
&& $A_{FB}(b)$  & $0.0984 \pm 0.0086 $ \lep & $0.096$ & \cr
&& $A_{FB}(c)$  & $ 0.090 \pm 0.019$ \lep & $ 0.068$ &\cr
&& $A_{LR} $  & $ 0.100 \pm 0.044$ \sld & $ 0.137$ & \cr
\noalign{\medskip}\noalign{\hrule}\noalign{\medskip}
&& $M_W$ (GeV)  & $79.91 \pm 0.39$ 
\ref\cdf{R. Abe {\sl et al.}, \prl{65}{90}{2243}.}\cdf & $80.18$ &\cr
&& $M_W/M_Z$ 	& $0.8798 \pm 0.0028 $ 
\ref\uatwo{J. Alitti {\sl et al.}, 
	\plb{276}{92}{354}.}\uatwo & 0.8793 & \cr
&& $\Gamma_W$  (GeV) & $2.12 \pm 0.11$ 
\ref\pdb{Particle Data Group, \prd{45}{92}{II}.}\pdb & 2.082 & \cr
\noalign{\medskip}\noalign{\hrule}\noalign{\medskip}
&& $g_L^2 $  & $0.3003 \pm 0.0039$ 
\paul & $0.3021$ & \cr
&& $g_R^2 $  & $0.0323\pm 0.0033 $ \paul & $0.0302$ & \cr
&& $g_A^e $  & $ -0.508 \pm 0.015 $ \paul & $-0.506$ & \cr
&& $g_V^e $  & $ -0.035\pm 0.017 $ \paul & $-0.037$ & \cr
&& $Q_W(Cs)$ & $-71.04 \pm 1.58 \pm [0.88]$ 
\ref\cesium{M.C. Noecker {\sl et al.}, 
	\prl{61}{88}{310}.}\cesium & $-73.20$ & \cr
\noalign{\medskip}\noalign{\hrule}\noalign{\smallskip}\noalign{\hrule}
}}$$
\centerline{{\bf TABLE VII}} 
\medskip
\noindent {\eightrm Experimental Values for Electroweak Observables
Included in Global Fit. The $Z^0$ measurements are the preliminary 1992 LEP
results taken from ref.~\lep. The couplings extracted from neutrino
scattering data are the current world averages taken from ref.~\paul. The
standard model values are for $m_t=150$ GeV and $M_{\sss H}=300$ GeV. We
have not shown theoretical errors in the standard model values due to
uncertainties in the radiative corrections, $\Delta r$, and due to errors
in $\mz$, as they are in general overwhelmed by the experimental errors. The
exception is the error due to uncertainty in $\alpha_s$, shown in square
brackets. We include this error in quadrature in our fits. The error in
square brackets for $Q_W(Cs)$ reflects 
the theoretical uncertainty in the atomic 
wavefunctions 
\ref\atomictheory{S.A. Blundell, W.R. Johnson, and J. Sapirstein,
\prl{65}{90}{1411}; V.A. Dzuba {\sl et al.}, \pla{141}{89}{147}.} 
\atomictheory\ 
and is also included in quadrature with the experimental
error.} 
\endinsert

As mentioned in the introduction, the same new physics responsible for the
anomalous TGV's will typically also contribute directly to the various
observables used in the fit. In our analysis we assume no cancellations
between these `direct' contributions and those due to the TGV's. Although
this may seem like a very strong assumption, we will see that in any case
the constraints obtained are rather weak.

We first consider the case in which only one of the TGV couplings, $\dkv$,
$\lv$ and $\dgz$, is nonzero at the scale $M$. In this case strong bounds
on this parameter may be obtained, since there is no possibility of
cancellations. Constraining one parameter at a time we obtain the following
values with 1$\sigma$ errors:
\eqa
\label\results
\dgz &= -0.033 \pm 0.031  \eolnn
\dkg &= 0.056 \pm 0.056  \eolnn
\dkz &= -0.0019 \pm 0.044 \eolnn
\lg &= -0.036 \pm 0.034 \eolnn
\lz &= 0.049 \pm 0.045 \eeol
\eeq
If taken at face value, these limits would imply that anomalous TGV's are
too small to be seen at LEP200.

Of course, although the bounds obtained in this way are the tightest bounds
that are possible, they are somewhat artificial. After all, real underlying
physics would produce more than just a single TGV. If we fit for all five
anomalous TGV's simultaneously, the constraints virtually disappear, due to
the possibility of cancellations. Several authors have fitted
$STU$-corrected expressions for observables to electroweak data, and have
concluded that the upper limit on these parameters is $O$(0.1-1). It is
clear that if one were to do fit a using the $STUVWX$-corrected expressions
displayed in Table V, then the limits on the six parameters would be looser
still. We find that at best, we can only conclude that the anomalous TGV
couplings are less than ${\cal O}(1)$. TGV's of this size would, of course,
be observable at LEP200.

The bounds given in eq.~\results\ are nevertheless interesting. These
values can be interpreted as an indication of the sensitivity of the global
fit of electroweak data to specific anomalous couplings. Once all of the
couplings are allowed to vary simultaneously, no significant bound remains.
This indicates that, in that part of the allowed region for which the TGV
couplings are large, cancellations occur among the contribution of the
various anomalous couplings to low-energy observables. Eq.~\results\ allows
one to gain a feel for the size of cancellations that would be required to
account for the low-energy data, should an anomalous TGV at the 10\% level
ever be discovered at LEP200.

Our results in this regard agree with those of ref.~\linearmad, who
similarly obtain no significant bound for these couplings, subject to the
somewhat stronger assumption that the effective theory be a linear
realization of the electroweak gauge group. As is discussed in this
reference, such a linear realization implies relationships among the
various TGV parameters, and so would be expected to lead to tighter
constraints than those obtained here. We here confirm this result within a
more general phenomenological analysis, without theoretical biases.

\section{Conclusions}

We have computed the bounds that may be obtained for CP-preserving
anomalous TGV's from current low-energy phenomenology. These bounds arise
due to the influence of these interactions, through loops, on well-measured
electroweak observables. We compute this influence using an
effective-lagrangian description, in which TGV interactions are imagined to
have been generated just below the scale for new physics, $\mu = M$, after
all of the hitherto undiscovered heavy particles have been integrated out.
Running this lagrangian, using the \msbar\ renormalization scheme, down to
the weak scale, $\mu \simeq 100$ GeV, then generates a collection of
secondary effective interactions. Unlike the TGV's, these new interactions
contribute directly to low-energy observables, and so their couplings may
be bounded by comparison to the data. We obtain limits on TGV's by
requiring that the contributions to these couplings due to their RG mixing
with the TGV's satisfy these experimental constraints. In so doing we are
tacitly assuming that no cancellations arise between the induced values for
these couplings, and their initial conditions at the new-physics scale,
$\mu = M$.

An interesting feature of this calculation is that the weak-scale effective
theory contains many higher-derivative interactions which contribute to
both the gauge-boson propagation, as well as to the fermion-boson vertex
corrections. We show how field redefinitions may be used to cast this
collection of terms into a simplified form, rendering  certain
cancellations among the anomalous TGV's explicit at the outset. One
approach involves putting all of the effects of new physics into the
gauge-boson self-energies. Another approach (the one followed in the
present article) involves recasting the effects of new physics into an
$STU$-sector plus a small set of higher-dimension fermion-boson vertex
terms. In either case, since the resulting `oblique' corrections
necessarily involve higher powers of the momentum-transfer, $q^2$, they may
{\it not} be parametrized purely in terms of the familiar variables $S$,
$T$ and $U$. The three additional parameters, $V$, $W$ and $X$, of
ref.~\stuvwx\ are also required. 

The necessity for these additional parameters may come as something of a
surprise, since we have assumed that the new physics scale, $M$, is large
in comparison to the weak scale, $\mz$. This would naively seem to justify
the neglect of the higher-derivative oblique corrections, since these
should be suppressed in their implications for low-energy observables by
factors of  $\mz^2/M^2$. The problem with this reasoning is that
loop-induced effects to $S$ through $U$ are at most of order $(g/4 \pi)^2$,
and this is the same size as $\mz^2/M^2$. As a result, TGV loop-induced
corrections to the $STU$ parameters are typically the same size as
higher-derivative oblique corrections, and so these must be properly
treated, as we have done.

We have found that the limits obtained in this way cannot in themselves
rule out TGV's that are large enough to be detectable once LEP runs at the
threshold for $W^\pm$ pair production. Couplings of this size {\it would}
have been ruled out if we had considered each TGV one at a time, with all
of the others constrained to be zero. In this case the data would constrain
the various TGV couplings to be 10\% or less. This shows how (fairly mild)
cancellations among the various TGV's would be required in low-energy
observables should these TGV's be directly detected at LEP200. Since the
underlying theory of new physics is unlikely to produce only one TGV at a
time, this also shows how misleading can be the practice of working with
TGV's one by one. A simultaneous fit of all of the TGV's to the data does
not yield useful bounds.

An interesting feature of the expressions in Table VI is the occurence of
some spectacular cancellations among anomalous TGV's. Although $S$, $T$ and
$U$ contain all of the combinations of the anomalous TGV's, such is not the
case for $V$, $W$ and $X$. In particular, the $\lambda_\ssv$ couplings do
not contribute to these parameters at all, $V$ depends only on $\dgz$, and
$X$ depends only on $\dkv$. This implies that each of the TGV couplings
tends to contribute only to particular kinds of observables at $q^2 =
\mz^2$ and $\mw^2$.

\bigskip
\centerline{\bf Acknowledgments}
\bigskip
S.G. and D.L. gratefully acknowledge helpful conversations and
communications with Paul Langacker. S.G. thanks Dean Karlen for useful
exchanges. We also thank R. Sinha for interesting communications. This
research was partially funded by funds from the N.S.E.R.C.\ of Canada, les
Fonds F.C.A.R.\ du Qu\'ebec, and by the Deutsche Forschungsgemeinshaft. 

\vfill\eject
\centerline{\bf Figure Captions}
\bigskip

\topic{Figure (1)}
The Feynman graphs through which the anomalous three- and four-point 
gauge-boson vertices contribute to the gauge boson vacuum polarization. 
The blobs represent anomalous couplings, and all other interactions are
standard. 

\topic{Figure (2)}
The Feynman graph through which the anomalous TGV's contribute to the gauge
boson-fermion vertex corrections. The blob represents the anomalous TGV,
and all other interactions are standard.

\listrefs

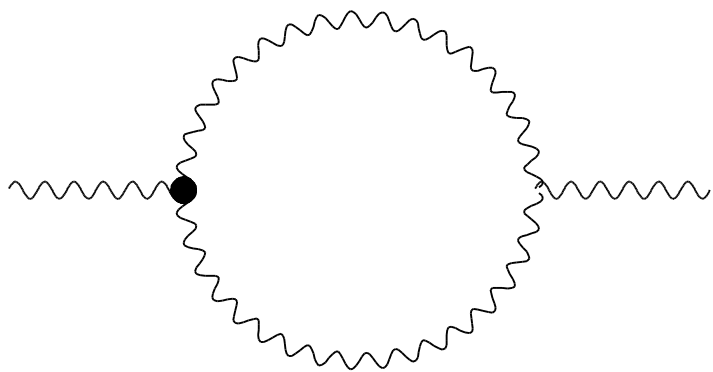
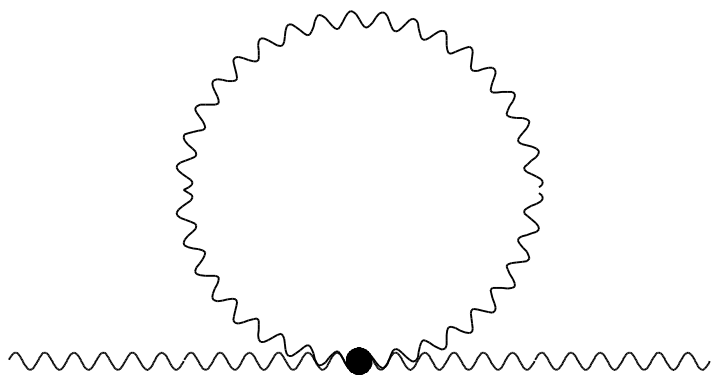
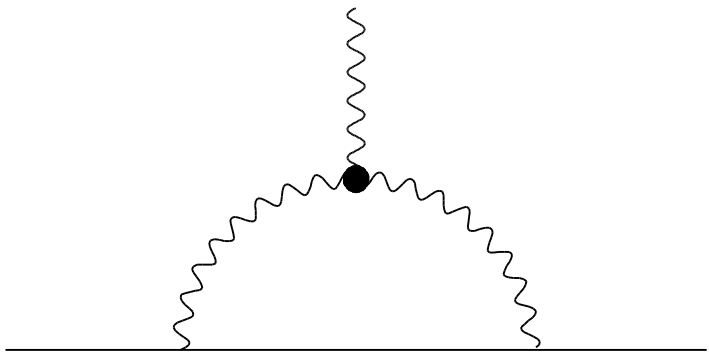
\bye